\documentclass[twocolumn]{aastex631}

\begin{document}

\title{The near infrared SED of young star clusters in the FEAST galaxies: Missing ingredients at 1 -- 5 $\mu$m}

\correspondingauthor{Alex Pedrini}
\email{alex.pedrini@astro.su.se}

\author[0000-0002-8222-8986]{Alex Pedrini}
\affiliation{Department of Astronomy, Oskar Klein center, Stockholm University, AlbaNova University center, SE-106 91 Stockholm, Sweden}

\author[0000-0002-8192-8091]{Angela Adamo}
\affiliation{Department of Astronomy, Oskar Klein center, Stockholm University, AlbaNova University center, SE-106 91 Stockholm, Sweden}

\author[0000-0001-8068-0891]{Arjan Bik}
\affiliation{Department of Astronomy, Oskar Klein center, Stockholm University, AlbaNova University center, SE-106 91 Stockholm, Sweden}

\author[0000-0002-5189-8004]{Daniela Calzetti}
\affiliation{Department of Astronomy, University of Massachusetts, 710 North Pleasant Street, Amherst, MA 01003, USA}

\author[0000-0002-1000-6081]{Sean T. Linden}
\affiliation{Steward Observatory, University of Arizona, 933 N Cherry Avenue, Tucson, AZ 85721, USA}

\author[0000-0003-4910-8939]{Benjamin Gregg}
\affiliation{Department of Astronomy, University of Massachusetts, 710 North Pleasant Street, Amherst, MA 01003, USA}

\author[0009-0008-4009-3391]{Varun Bajaj}
\affiliation{Space Telescope Science Institute, 3700 San Martin Drive, Baltimore, MD 21218, USA}

\author[0000-0002-2918-7417]{Jenna E. Ryon}
\affiliation{Space Telescope Science Institute, 3700 San Martin Drive, Baltimore, MD 21218, USA}

\author[0000-0002-1832-8229]{Anne S. M. Buckner}
\affiliation{Cardiff Hub for Astrophysics Research and Technology (CHART), School of Physics \& Astronomy, Cardiff University, The Parade, CF24 3AA Cardiff, UK}

\author[0009-0003-6182-8928]{Giacomo Bortolini}
\affiliation{Department of Astronomy, Oskar Klein center, Stockholm University, AlbaNova University center, SE-106 91 Stockholm, Sweden}

\author[0000-0001-6291-6813]{Michele Cignoni}
\affiliation{Dipartimento di Fisica, Università di Pisa, Largo Bruno Pontecorvo 3, 56127, Pisa, Italy}
\affiliation{INAF - Osservatorio di Astrofisica e Scienza dello Spazio di Bologna, via Piero Gobetti 93/3, 40129 Bologna, Italy}
\affiliation{INFN, Largo B. Pontecorvo 3, 56127, Pisa, Italy}

\author[0000-0001-6464-3257]{Matteo Correnti}
\affiliation{INAF Osservatorio Astronomico di Roma, Via Frascati 33, 00078, Monteporzio Catone, Rome, Italy}
\affiliation{ASI-Space Science Data Center, Via del Politecnico, I-00133, Rome, Italy}

\author[0000-0002-5259-4774]{Ana Duarte-Cabral}
\affiliation{Cardiff Hub for Astrophysics Research and Technology (CHART), School of Physics \& Astronomy, Cardiff University, The Parade, CF24 3AA Cardiff, UK}

\author[0000-0002-1723-6330]{Bruce~G.~Elmegreen}\affiliation{Katonah, NY 10536, USA}

\author[0000-0002-2199-0977]{Helena Faustino Vieira}
\affiliation{Department of Astronomy, Oskar Klein center, Stockholm University, AlbaNova University center, SE-106 91 Stockholm, Sweden}

\author[0000-0001-8608-0408]{John S. Gallagher}
\affiliation{Department of Astronomy, University of Wisconsin-Madison, 475 N. Charter Street, Madison, WI 53706, USA}

\author[0000-0002-3247-5321]{Kathryn~Grasha}
\altaffiliation{ARC DECRA Fellow}
\affiliation{Research School of Astronomy and Astrophysics, Australian National University, Canberra, ACT 2611, Australia}   
\affiliation{ARC Centre of Excellence for All Sky Astrophysics in 3 Dimensions (ASTRO 3D), Australia}

\author[0000-0001-8348-2671]{Kelsey E. Johnson}
\affiliation{Department of Astronomy, University of Virginia, Charlottesville, VA 22904, USA}

\author[0000-0003-3893-854X]{Mark R. Krumholz}
\affiliation{Research School of Astronomy \& Astrophysics, Australian National University, 233 Mt Stromlo Rd, Stromlo, ACT 2611, Australia}

\author[0009-0009-5509-4706]{Drew
 Lapeer}
\affiliation{Department
 of Astronomy, University of Massachusetts, 710 North Pleasant Street, Amherst, MA 01003, USA}

\author[0000-0001-8490-6632]{Thomas S.-Y. Lai}
\affil{IPAC, California Institute of Technology, 1200 E. California Blvd., Pasadena, CA 91125}

\author[0000-0003-1427-2456]{Matteo Messa}
\affiliation{INAF – OAS, Osservatorio di Astrofisica e Scienza dello Spazio di Bologna, via Gobetti 93/3, I-40129 Bologna, Italy}

\author[0000-0002-3005-1349]{Göran Östlin}
\affiliation{Department of Astronomy, Oskar Klein center, Stockholm University, AlbaNova University center, SE-106 91 Stockholm, Sweden}

\author[0009-0002-2481-2217]{Linn Roos}
\affiliation{Department of Physics and Astronomy, Uppsala University, Box 516, SE-751 20 Uppsala, Sweden}

\author[0000-0002-0806-168X]{Linda J. Smith}
\affiliation{Space Telescope Science Institute, 3700 San Martin Drive, Baltimore, MD 21218, USA}

\author[0000-0002-0986-4759]{Monica Tosi}
\affiliation{INAF - Osservatorio di Astrofisica e Scienza dello Spazio di Bologna, via Piero Gobetti 93/3, 40129 Bologna, Italy}



\begin{abstract}

We present a combined HST and JWST 0.2--to--5 $\mu$m analysis of the spectral energy distributions (SEDs) of emerging young star clusters (eYSCs) in four nearby galaxies from the Feedback in Emerging extrAgalactic Star clusTers (FEAST) survey: M51, M83, NGC~628, and NGC~4449. These clusters, selected for their bright Pa$\alpha$ and 3.3 $\mu$m polycyclic aromatic hydrocarbon (PAH) emission, are still associated to their natal gas cloud and have been largely missed in previous HST optical campaigns.
We modeled their SEDs using the CIGALE fitting code and identified: i) a systematic flux excess at 1.5–2.5 $\mu$m that is not accounted for by current stellar population models; ii) the preference for a set of dust model parameters that is not aligned with expectations from self-consistent analyses of star-forming regions, suggesting model shortcomings also in the 3--5 $\mu$m. The near-infrared (NIR) excess is most prominent in low-mass ($\leq 3000$ M$_\odot$) and young ($\leq 6$ Myr) clusters. Additionally, we see that the SED fitting analysis wrongly assigns ages  $\geq6$ Myr to a fraction of strong Pa$\alpha$ emitters with equivalent widths suggestive of significantly younger ages. A parallel analysis with the {\tt slug} code suggests that stochastic initial mass function (IMF) sampling of pre-main-sequence stars combined with extinction might partially reduce the gap.  We conclude that the inclusion of young stellar object SEDs, along with more realistic sampling of the cluster IMF, might be needed to fully account for the stellar population and dust properties of eYSCs.  


\end{abstract}

\keywords{Star clusters (1567), Polycyclic aromatic hydrocarbons (1280), Star formation (1569), Spectral energy distribution (2129), Galaxies (573)}


\section{Introduction} \label{sec:intro}
The appearance and evolution of galaxies are largely governed by their star formation activity.
Due to its hierarchical nature \citep{Elmegreen96}, star formation is linked to the presence of clustered structures. At parsec scales, thousands of young star clusters (YSCs) contribute to the observed UV light of their host galaxies and, via feedback processes, regulate the star-formation cycle in the galaxy \citep[][and references therein]{Adamo20}. 
While high resolution studies of star clusters in the Milky Way allow us to dissect single clusters in their stellar, gas, and dust components \citep[e.g.,][]{Zucker23}, they suffer from line-of-sight confusion and dust obscuration, which prevent the collection of large, unbiased samples of the cluster population. Probing the entire star cluster population in nearby galaxies is fundamental to understanding star formation as a function of galactic environment and physical properties. 

During the last three decades, the Hubble Space Telescope (HST) has paved the way for an unprecedented census of YSCs in nearby galaxies \citep[][among many others]{Whitmore11,Chandar14,Johnson16,Adamo17,Messa18a,Turner22,Maschmann24}. However, the observed demographics of YSCs, as well as the physical processes behind cluster formation and evolution \citep[e.g.,][]{Kruijssen15,Krumholz19}, cannot be fully understood without grasping the very young stage of life of YSCs i.e., the emerging phase.  
In the emerging phase, YSCs are still embedded in their dusty natal cloud and stellar feedback \citep[predominantly photoionization and radiation pressure,][]{Johnson03,Johnson15,Kim18,Krumholz19,Menon22,Dobbs23,Pathak25} from massive stars dramatically impacts the surroundings, shaping a multi-phase interstellar medium (ISM). This stellar feedback is not instantaneous, and thus for a certain time emerging YSCs (eYSCs) are invisible at optical wavelengths and largely missed from the aforementioned optical surveys. \cite{Messa21} estimated that up to 60\% of eYSCs are missed for ages between 1 and 5 Myr using NUV-optical-NIR HST observations in NGC 1313. 

Recently, the advent of JWST has enabled us to detect most, if not all, eYSC populations in nearby galaxies, outside of the Local Group. Thanks to its unprecedented sensitivity and spatial resolution, several campaigns have been characterizing eYSCs and their surrounding star-forming regions in nearby galaxies using JWST NIRCam bands and NIRSpec spectroscopic data \citep[e.g.,][]{Whitmore23,Rodriguez23,Levy24,Linden24,Pedrini24,Sun24,Whitmore25,Knutas25}. These authors show that eYSCs present an excess of 3.3 $\mu$m polycyclic aromatic hydrocarbons (PAH) emission from their photo-dissociation regions (PDRs, \citealt{Hollenbach99}) and they are able to clear their surrounding envelopes due to stellar feedback in a few Myr.  
However, most of the results presented in these studies rely on correct estimations of eYSC physical properties. Specifically, we need to carefully account for a detailed NIR spectral modeling of emission from star-forming regions in order to obtain accurate derivations of eYSCs ages and stellar masses. In optical surveys, single stellar population (SSP) synthesis models that account for stellar and nebular contributions \citep[e.g., {\tt yggdrasil}][]{Zackrisson11} have been extensively leveraged and can successfully describe the spectral energy distribution (SED) of YSCs with ages down to $\sim 5$ Myr, where the results become extremely sensitive to the most massive stars included in the model, and the way those massive stars form and evolve \citep{Wofford16}. In this respect, due to pre-JWST low sensitivity and resolution of NIR observations in nearby galaxies, models for NIR stellar isochrones have been less explored/tested. Moreover, the NIR is not only sensitive to ionized gas and stellar continuum but also to the complex emission of dust around stars and in PDRs \citep{Indebetouw06,Peeters23,Chown24,Schroetter24}. 
With JWST, we can now resolve individual extragalactic star clusters, their associated compact HII region and warm dust in proximity of the newly formed stars within the Local Volume ($d < 11$ Mpc) at parsec-scale resolution. 

The Feedback in Emerging extrAgalactic Star clusTers (FEAST, GO 1783, PI A.Adamo) survey plays a key role in advancing the observational interpretation of extragalactic eYSC at parsec scales. FEAST encompasses JWST/NIRCam observations of six nearby galaxies spanning a wide range of physical properties. Moreover, these observations were designed to both sample large fields of view (covering most of the targets) and achieve high resolution, allowing for statistical studies of individual eYSCs within single galaxies (Adamo et al., in prep., \citealt{Gregg24}). 
This survey aims to investigate the primary physical mechanisms driving the emergence of eYSCs by measuring levels of obscuration across clusters at different evolutionary stages, and exploring how these trends correlate with galactic environment. \cite{Gregg24} analyzed the relation between star formation rate and 3.3 $\mu$m PAH emission for eYSCs in NGC~628, finding that age variations and physical properties of PAHs might have strong impacts on the linearity of this relation. In Gregg et al.~(subm.), the authors have explored variations in this relation as a function of galactic physical properties across the diverse FEAST sample of galaxies. 
Furthermore, \cite{Pedrini24} illustrated that eYSCs in NGC~628 are linked to star-forming regions where HII regions and PDRs traced by 3.3 $\mu$m PAH emission are still compact, and their morphology is constrained by cluster evolution. \cite{Knutas25} have studied the eYSC population in M83, with a specific focus on the emerging timescales of eYSCs in different environments within the galaxy disk as well as cluster mass. 

With these unprecedented results on the demography of eYSCs in nearby galaxies, comes the need for comprehensive models that accurately describe the SED of eYSCs. In this context, the excellent sensitivity provided by JWST has revealed populations of embedded clusters that present stellar masses ranging from tens to tens of thousands of M$_{\odot}$ (e.g., \citealt{Knutas25}, Linden et al.,~in prep.). In this low-mass regime, stochastic effects due to initial mass function (IMF) sampling are no longer negligible, and a full consideration of these effects becomes necessary \citep{Maiz09, Calzetti10}. Under these conditions, distinguishing the main sources of emission from eYSCs in the NIR window becomes crucial for a more complete understanding and interpretation of their NIR SED.  

In this paper, we use JWST/NIRCam observations and ancillary HST data to present a multi-wavelengths analysis of the UV-optical-NIR SED of eYSCs in M51, M83, NGC~628 and NGC~4449, which are part of the FEAST survey. In particular, we derive physical properties of eYSCs using state-of-the-art SED fitting codes and discuss their performances, with a specific focus on the NIR wavelengths. After an introduction of the observations and the eYSC populations (Sec.~\ref{sec:obs}), we present the adopted SED fitting methodology (Sec.~\ref{sec:CIGgrid}). The results presented in this paper (Sec.~\ref{sec:CIGres}) reveal discrepancies between the NIR-recovered best fluxes from models and our observations (Sec.~\ref{sec:NIRexcess}). We discuss possible explanations in details and various interpretations of this issue, focusing on the analysis of the impact of stochastic IMF sampling effects (Sec.~\ref{sec:discussion}). The paper ends with our conclusions and summary of findings in Sec.~\ref{sec:conclusion}.

\section{Imaging data} \label{sec:obs}

\begin{deluxetable*}{lcccccc}[htb!]
\tablecaption{Summary of the FEAST galaxies analysed in this paper. From left to right: distance, stellar mass, UV extinction corrected star formation rate (SFR), central metallicity from HII regions oxygen abundance, foreground Milky Way attenuation and number of eYSCI presented in this work (see Sec.~\ref{sec:eyscPresentation}).
Reference dictionary: 1 - \cite{Calzetti15}, 2 - \cite{Leroy21}, 3 - \cite{Berg20}, 4 - \cite{Bresolin16}, 5 - \cite{Sabbi18}, 6 - \cite{Tully09}, 7 - \cite{Csornyei23}, 8 - \cite{Tully13}, 9 - \cite{Schlafly11}, 10 - \cite{Pilyugin15}. \label{tab:targets}}
\tablenum{1}
\tablehead{\colhead{Galaxy} & \colhead{$d$~[Mpc]} & \colhead{$\log(M_*$/ M$_{\odot}$)} & \colhead{SFR~[M$_{\odot}/$yr]} & \colhead{12 + $\log({\rm O}/{\rm H})$} & \colhead{Foreground A$_V$} & \colhead{\# eYSCI}}
\startdata
M51 & 7.5 (7) & 10.38 (1) & 6.88 (1) & 8.65 (3) & 0.095 (9) & 1707  \\
M83 & 4.7 (8) & 10.53 (2) & 4.17 (2) & 8.9 (4) & 0.182 (9) & 1087 \\
NGC~628 & 9.84 (6) & 10.34 (2) & 1.74 (2) & 8.51 (3) & 0.192 (9) & 784 \\
NGC~4449 & 4.0 (5) & 9.04 (1) & 0.94 (1) & 8.26 (10) & 0.053 (9) & 249 \\
\enddata
\end{deluxetable*}

\subsection{FEAST galaxies}
For this paper, we use the observations of four galaxies, part of the FEAST program: M51, M83, NGC~628 and NGC4449. The properties of these galaxies are summarized in Tab.~\ref{tab:targets}. JWST/NIRCam observations of M51, M83, NGC~628 and NGC4449 have been obtained between January 2023 and June 2024 as part of the FEAST program. For each galaxy, we obtained simultaneous observations in short wavelengths (F115W, F150W, F187N, F200W) and long wavelengths channels (F300M, F335M, F405N, F444W). In NGC~628, the F277W filter replaced the F300M. We exploited a FULLBOX 4TIGHT dither pattern to ensure high quality sampling in large field of views (e.g., $2' \times6'$ in NGC~628) and of the point spread functions (PSFs). 
We refer to Adamo et al.,~(in prep.) for an in-depth description of the data reduction, while here we present an overview of the main steps. Our NIRCam observations have been downloaded from the Mikulski Archive for Space Telescopes (MAST). We reduced the data using the distributed pipeline (version 1.12.5) with calibration data context number 1169.

For these galaxies, we download additional available HST archival observations covering the UV-optical range from the MAST archive and we reduced them using standard data-reduction steps (see Adamo et al.,~in prep.). Additional information on these ancillary data is found in Table~\ref{tab:filters}, which presents an overview of the filters employed in this analysis across the sample of galaxies. 
Finally, we resampled both NIRCam and HST mosaics to a common pixel grid of 0.04$''/$pix and we aligned them to the same reference system using GAIA \citep{Gaia23}.

\begin{deluxetable*}{lcc}
\tablecaption{Overview of the observational data. HST/WFC3 observations are from \#13340 (Van Dyk), \#12762 (Kuntz), \#13364 (Calzetti), \#17225 (Calzetti), \#11360 (O'Connell) and \#12513 (Blair). HST/ACS observations are from \#10452 (Beckwith), \#9796 (Miller), \#10402 (Chandar) and \#10585 (Aloisi). \label{tab:filters}}
\tablenum{2}
\tablehead{\colhead{Galaxy} & \colhead{JWST/NIRCam} & \colhead{HST} }
\startdata
M51 & F115W, F150W, F187N, F200W, & WFC3/F275W, WFC3/F336W, ACS/F435W, ACS/F555W, \\  & F300M, F335M, F405N, F444W &  ACS/F658N, WFC3/F689M, ACS/F814W \\
\\
M83 & F115W, F150W, F187N, F200W, & WFC3/F225W, WFC3/F275W, WFC3/F336W, WFC3/F438W, WFC3/F547M, \\  & F300M, F335M, F405N, F444W & WFC3/F555W, WFC3/F657N, WFC3/F689M, WFC3/F814W \\
\\
NGC~628 & F115W, F150W, F187N, F200W, & WFC3/F275W, WFC3/F336W, ACS/F435W, ACS/F555W, \\  & F277W, F335M, F405N, F444W & ACS/F658N, ACS/F814W \\
\\
NGC~4449 & F115W, F150W, F187N, F200W, & WFC3/F275W, WFC3/F336W, ACS/F435W, ACS/F555W,\\  & F300M, F335M, F405N, F444W & ACS/F658N, ACS/F814W \\
\enddata
\end{deluxetable*}

\subsection{eYSC catalogues} \label{sec:eyscPresentation}

Stellar feedback mechanisms in eYSCs produce HII regions and PDRs that can be directly traced with multiple JWST/NIRCam filters. In particular, the F187N and F405N filters trace emission from the hydrogen recombination lines Pa$\alpha$ and Br$\alpha$, respectively. On the other hand, the 3.3 $\mu$m PAH feature observed by means of the F335M filter represents an optimal tracer for PDRs in star-forming regions \citep[e.g.,][]{Peeters23,Chown24}. For these reasons, we based our selection of eYSCs on their photometric properties in the aforementioned filters. We present and describe the full methodology to detect and classify eYSCs in Adamo et al.~(in prep.). Here we summarize the main steps.
We identified eYSCs as peaked emission in the NIR hydrogen recombination lines (Pa$\alpha$-1.87$\mu$m and Br$\alpha$-4.05$\mu$m) and in the 3.3$\mu$m PAH feature. As already described in \cite{Gregg24} and \cite{Pedrini24}, we refer to this eYSC class as eYSCI, while the eYSCII class represents clusters without compact 3.3$\mu$m PAH emission. 

To obtain Pa$\alpha$, Br$\alpha$ and 3.3 $\mu$m PAH maps of our sample of galaxies, we removed the stellar and dust continuum from filters containing the emission features. We subtract the stellar continuum from the F187N filter using the adjacent broad filters F150W and F200W. Since the F200W filter also contains Pa$\alpha$ emission, we applied an iterative procedure to remove its contribution, as described in \cite{Gregg24} and \cite{Calzetti24}.
On the other hand, for the F335M and F405N filters, we removed the stellar and dust continuum using the F300M (F277W in NGC~628) and F444W filters. In the case of F405N, the subtraction was also performed iteratively, as the F444W filter contains Br$\alpha$ emission. This continuum subtraction procedure is fully described in \cite{Gregg24} and results in continuum-subtracted emission line maps for Pa$\alpha$, Br$\alpha$ and the 3.3 $\mu$m PAH feature.

We present now the identification of eYSCI clusters for M51, M83, NGC~628 and NGC~4449. The candidates are independently identified on the three emission maps using \texttt{SEP}, a python package for source extraction \citep{Barbary16}, resulting in three different catalogs, one for each emission line map. We then visually inspected all the extracted sources with the aim of re-centering and cleaning contaminants in each reference emission map. 
The catalogs of visually inspected sources are then used to perform aperture photometry in all available JWST and HST sets of filters (see Tab.~\ref{tab:filters}). For the closer galaxies, M83 and NGC~4449, we adopted an aperture radius of 5 pixels, which corresponds to physical sizes of $\sim$4.55 and $\sim$3.88 pc, respectively. In the case of M51 and NGC~628, we used a radius of 4 pixels ($\sim$5.8 pc and $\sim$7.8, respectively). To correct for background emission, we subtract the mode value estimated from an annulus aperture, centered in the source position, with a radius of 7 pixels for both M83 and NGC~4449 and 6 pixels for M51 and NGC~628. We adopted 2 pixels for the width of the annulus aperture. 
Furthermore, we used concentration index-based aperture corrections to account for different point spread functions (PSFs) within the UV-optical-NIR windows (see Adamo et al.,~in prep. and \citealt{Knutas25}).

To obtain the final science-ready catalog of eYSC candidates, we corrected the photometry for Milky Way $A_{\rm V}$ extinction along the line of sight (Tab.~\ref{tab:targets}). 
For each emission map, we produced a photometric catalog containing positions of the eYSC candidates and their flux densities (in mJy and ABmag). To ensure clear detections, we additionally required a magnitude error $<0.3$ in the F187N and F200W filters for sources extracted from the Pa$\alpha$, and in the F405N (F335M) and F444W filters for sources extracted within the Br$\alpha$ (3.3 $\mu$m PAH) map.

As a final step, we matched the three catalogs by their positions to determine their class: eYSCI are detected in Pa$\alpha$ and 3.3 $\mu$m PAH (Br$\alpha$, due to the lower resolution, is included but not considered a necessary condition); eYSCIIs are detected in Pa$\alpha$ (and possibly Br$\alpha$); 3.3 $\mu$m PAH peak sources do not have a clear peaked emission in Pa$\alpha$ and/or Br$\alpha$. We adopted a 4 pixel separation between emission peak centers as our criterion for catalog matching. To test the robustness of this choice, we also performed the matching using 5 and 6 pixel separations, finding that it introduces only a $\sim$2\% difference in the number of recovered eYSCI \citep[see also][]{Knutas25}.

\begin{figure*}[htb]
    \centering
    \includegraphics[width = 1.0\textwidth]{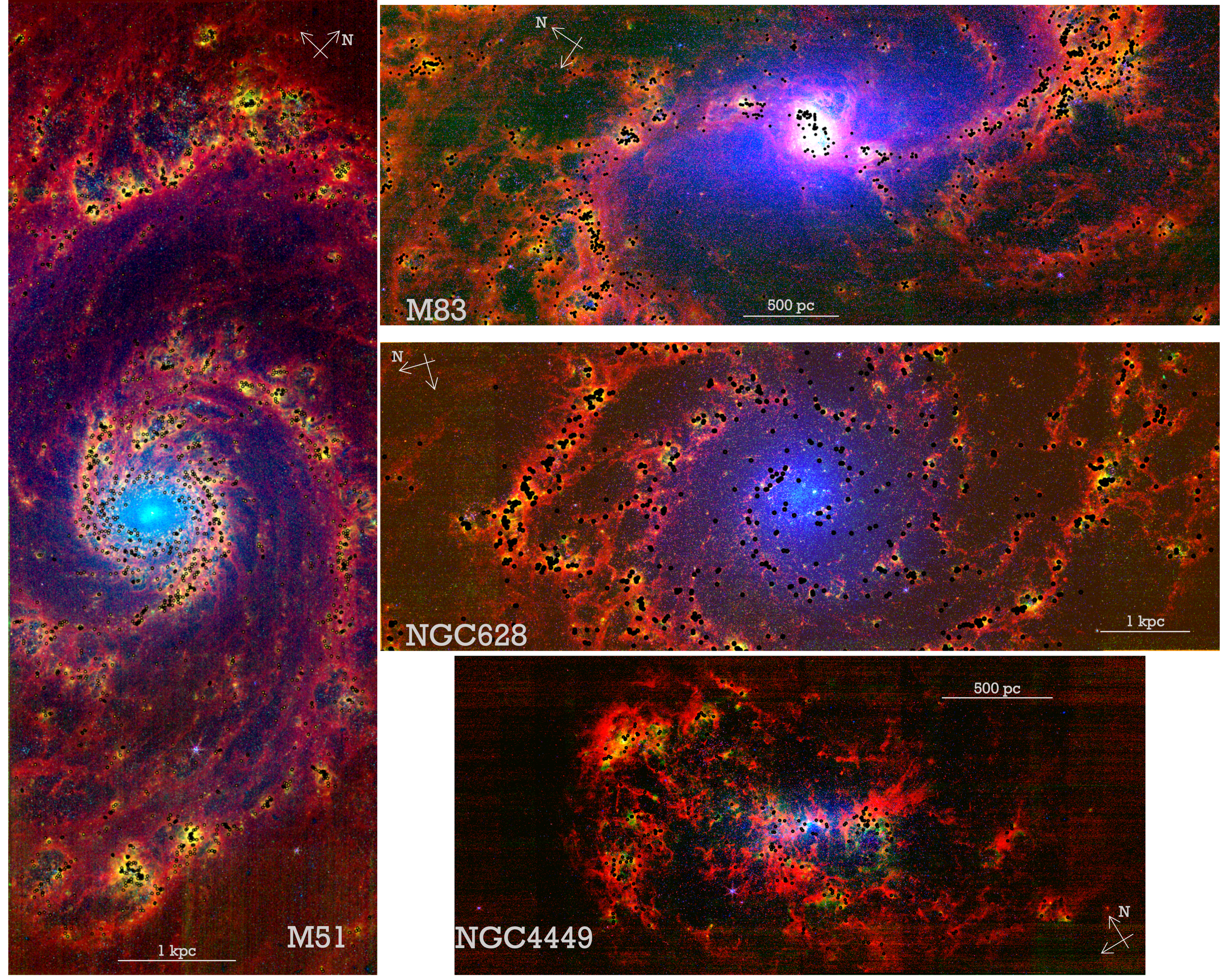}
    \caption{NIRCam detected eYSC populations in M51 (left panel), M83 (top right panel), NGC~628 (middle right panel) and NGC~4449 (bottom right panel). For each galaxy, the RGB color scheme is the following: F335M continuum subtracted map (red), F187N continuum subtracted map (green), F115W map (blue). For NGC~628, we replaced F187N with F405N, due to better visual quality of the map. We highlight the position of the eYSC populations with black dots. In the three spiral galaxies, eYSCs are located along the spiral arms, where they are associated with the highest concentration of star-forming regions in the galactic disks.}
    \label{fig:rgb_collage}
\end{figure*}

For the scientific goal of this paper, we consider only eYSCI with secure detection (S/N $> 3$) in all 8 NIRCam bands. The presence of ionized gas and PAH emission indicates that these objects are young and likely dusty, making this sample ideal for studying cluster SEDs in the NIR wavelengths. This last step leads to the final eYSCI catalogs, which contain 1707, 1087, 784 and 249 objects for M51, M83, NGC~628 and NGC~4449, respectively (last column of Tab.~\ref{tab:targets}). We present the eYSCI populations in the JWST field of view within the four targets in Fig.~\ref{fig:rgb_collage}, where the position of each cluster is represented by a black dot. 
We see a strong association between the eYSCI populations and the spiral structures of M51, NGC~628, M83, and NGC4449 (although irregular), where we observe enhanced ionized gas emission (Pa$\alpha$, green channel) and PAH emission from the PDRs (red channel) due to intense star formation activity.  

\subsection{Optical YSC catalogs} \label{sec:opticalCat}
Using HST imaging, we also produced optically identified YSC catalogs for our sample of galaxies. For all the galaxies, the cluster identification is based on compact sources emitting in the HST/F555W filter. In M51, the initial catalog used here relies on the LEGUS \citep[Legacy ExtraGalactic UV Survey,][]{Calzetti15} catalog presented in \cite{Messa18a}. Within this catalog, we selected cluster with an assigned morphological class of 1, 2 or 3 (see \citealt{Adamo17} for a description of these classes). Moreover, we required these clusters to lie within the JWST field of view, obtaining a final catalog consisting of 2361 YSCs. We adopted a similar procedure for the NGC~4449 LEGUS catalog \citep{Whitmore20}, obtaining 573 optically identified YSCs. For M83, we used the optical YSC catalog presented in \cite{Knutas25} and we applied analogous masks and selections as in M51 and NGC~4449, resulting in 3121 cluster candidates. We applied the same method for the optical YSC population in NGC~628 (presented in Adamo et al.,~in prep.), obtaining 3784 optical YSCs. We note that, on average across the four galaxies, 4\% of the optically selected YSCs coincide with the position of an eYSC.

The positions of the cluster candidates (centered using HST/F555W) are used to perform aperture photometry in all the HST and JWST available filters. Aperture photometry, local sky background and aperture correction are performed by matching the same physical scales in all HST and JWST filters using the same settings as described for the eYSC catalogs in Sec.~\ref{sec:eyscPresentation}. We will discuss the results obtained with these catalogs, applying the appropriate S/N and age selections, in Sec.~\ref{sec:discussion}.

\section{CIGALE modelling} \label{sec:CIGgrid}

To analyze the 0.2--to--5 $\mu$m SEDs of the observed YSCs and eYSCs and to constrain their physical properties (e.g., age, mass, attenuation), we performed SED fitting using the Code Investigating GALaxy Emission \citep[CIGALE,][]{Boquien19}. 
Although this code is originally designed to fit the SED of galaxies, it has been effectively exploited in the study of star cluster light \citep{Fensch19,Turner21,Pedrini24}. Linden et al.~(in prep.) presents a detailed complementary analysis showing a performance comparison between CIGALE and an alternative model widely used in cluster analysis \citep[{\tt yggdrasil},][]{Zackrisson11} in recovering physical properties of YSCs and eYSCs in the spiral galaxy NGC~628.

 \subsection{CIGALE grid for eYSCs}
CIGALE works in two main steps. The first step consists of computing a grid of models that encompasses the generation of stellar, nebular and dust emission models covering a pre-defined parameter space. Subsequently, in the second step the code performs a Bayesian-like $\chi^2$ analysis to fit the model grid to the observed cluster SEDs \citep{Yang20}.
As a result, CIGALE outputs the best values for the input-defined parameter space, as well as best fit fluxes for each filter. For each galaxy, we used the final eYSCI catalogs as input observations files. For the HST filters, we flagged the observed flux in a given filter as an upper limit if the S/N ratio is lower than 3.

As input for the CIGALE models, we use the parameter grid designed in Linden et al.~(in prep.), shown in Tab.~\ref{tab:gridCIGALE}. We consider an eYSC SED to be generated by an instantaneous burst of star formation, which consists of a decaying exponential function of timescale 0.001 Myr. As clusters are identified as bright sources in Pa$\alpha$, we limit the age range of the stellar population to vary between 1 and 10 Myr with steps of 1 Myr. The stellar emission is modeled from the \cite{Bruzual03} stellar models, considering a \cite{Chabrier03} IMF and solar metallicity (40\% solar for NGC~4449). Nebular emission models are based on CLOUDY grids \citep{Inoue11,Ferland13,Boquien19}. 
To account for dust attenuation, we use the modified starburst model, which allows the user to parametrize the starburst attenuation law from \cite{Calzetti00}. Additionally, the \texttt{E\_{\tt BV}\_{\tt factor}} parameter (Tab.~\ref{tab:gridCIGALE}) regulates the reduction factor between the attenuation computed for the emission lines and the stellar continuum attenuation. For the emission lines, we adopted a Milky Way extinction curve \citep[see][for details in the attenuation model]{Boquien19}. Finally, we use the dust emission templates from \cite{Draine14}.

\begin{deluxetable*}{lll}
\tablecaption{CIGALE adopted grid, as described in Sec.~\ref{sec:CIGgrid}. Fixed parameter are marked with *. \label{tab:gridCIGALE}}


\tablenum{3}

\tablehead{\colhead{Parameter} & \colhead{Type} & \colhead{Values} }

\startdata
{\texttt{tau\_{\tt main}}}* &  SFH & 0.001 Myr  \\
\texttt{age} & SFH  & 1 to 10 Myr (step 1 Myr)   \\
{\texttt{tau\_{\tt burst}}}* & SFH & 0.001 Myr  \\
\texttt{burst\_{age}}* & SFH & 1 Myr  \\
\texttt{f\_{burst}}* & SFH & 0  \\
\texttt{sfr\_{0}}* & SFH & 1.0  \\
\texttt{normalise}* & SFH & True  \\
\texttt{imf} & Stellar & \cite{Chabrier03}  \\
\texttt{metallicity}* & Stellar & 0.02 (0.008 for NGC~4449) \\
\texttt{separation\_{\tt age}}* & Stellar & 10 Myr  \\
\texttt{logU} & Nebular & -3.5, -3.0, -2.5, -2.0  \\
\texttt{zgas}* & Nebular & 0.02 (0.008 for NGC~4449) \\
\texttt{ne} & Nebular & 10, 100 cm$^{-3}$ \\
\texttt{f\_{\tt esc}} & Nebular & 0.01, 0.1, 0.2, 0.3, 0.4, 0.5, 0.6 \\
\texttt{f\_{\tt dust}} & Nebular & 0.01, 0.1, 0.2, 0.3 \\
\texttt{lines\_{\tt width}}* & Nebular & 50 km/s \\
\texttt{E\_{\tt BV}\_{\tt lines}} & Attenuation & 0.01, 0.1 to 5.0 (step 0.1) \\
\texttt{E\_{\tt BV}\_{\tt factor}} & Attenuation & 0.4, 0.5, 0.6, 0.8, 1.0 \\
\texttt{uv\_{\tt bump}\_{\tt wavelength}}* & Attenuation & 217.5 nm \\
\texttt{uv\_{\tt bump}\_{\tt width}}* & Attenuation & 35.0 nm \\
\texttt{uv\_{\tt bump}\_{\tt amplitude}}* & Attenuation & 0.0 nm \\
\texttt{Ext\_{\tt law}\_{\tt emission}\_{\tt lines}} & Attenuation & Milky Way \citep{Cardelli89} \\
\texttt{Rv}* & Attenuation & 3.1 \\
\texttt{qpah} & Dust & 0.47, 1.12, 1.77, 2.50, 3.19, 3.90, 4.58 \\
\texttt{umin} & Dust & 0.3, 0.5, 0.7, 1.0 {\it informed grid: 1,10 } \\
\texttt{alpha}* & Dust & 2.0 \\
\texttt{gamma} & Dust & 0.01, 0.05, 0.1, 0.5 {\it informed grid: 0.5,1 } \\
\enddata

\end{deluxetable*}

\subsubsection{Testing parameter choice in CIGALE}\label{sec:dustgrid}

CIGALE uses several parameters to control the dust and ionized gas emission, as introduced in the previous section and presented in Tab.~\ref{tab:gridCIGALE}. In general, we observe that changing the parameter grid for nebular gas emission ($\log(U)$, n$_e$, f$_{\rm esc}$, f$_{\rm dust}$) does not change the recovered physical parameters or the goodness of the fit ($\chi^2$) in a noticeable way. However, the inclusion of the nebular continuum and emission in the models is fundamental to fit the fluxes in the narrow bands centered on H$\alpha$, Pa$\alpha$, Br$\alpha$. In Appendix~\ref{app:opt4449}, we show the impact of including/excluding the narrowband filters from the SED of the optical YSCs in NGC~4449. As already shown by \cite{Whitmore20}, the inclusion of at least one narrowband transmitting a hydrogen emission line (e.g., H$\alpha$) mitigates the age-extinction degeneracy.
We compare the 0.2--to--5~$\mu$m FEAST SED outputs including narrowbands with the previous \cite{Whitmore20} output showing a very good agreement with the analysis performed by previous teams. We conclude that as long as at least one nebular line is included in the cluster SED fit, then the results are relatively insensitive to the nebular grid choice. 

\begin{figure*}[htb]
    \centering
    \includegraphics[width = 0.80\textwidth]{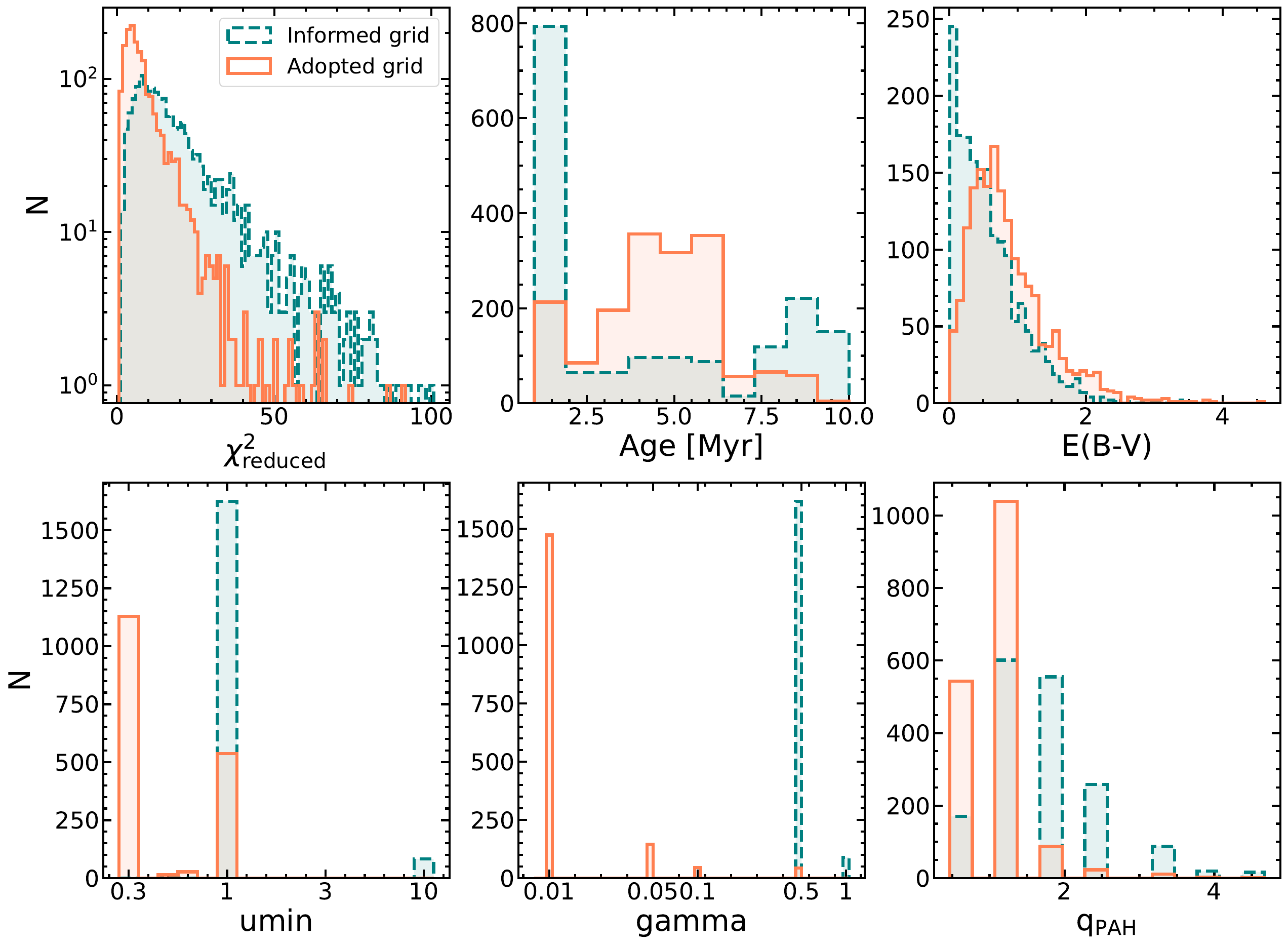}
    \caption{Recovered $\chi^2_{\rm reduced}$, Age, E(B-V), $U_{\rm min}$, $\gamma$ and q$_{\rm PAH}$ distributions from CIGALE for two different grids in M51. While the orange histograms show the results for the grid adopted in this paper (see Tab.~\ref{tab:gridCIGALE}), the blue distributions result from the empirically motivated dust grid (IF grid), as described in the text. }
    \label{fig:comparison_infgrid}
\end{figure*}

As for the dust parameters, we noticed a significant impact on the resulting recovered cluster physical properties. Taken at face value, we do not have enough information in our SEDs to derive stringent constraints on the dust. The dust parameters require observational constraints in the MIR-FIR wavelength range, which falls outside our observations and cannot be obtained at the physical scales studied here. Parameters linked to dust heating and dust mass heated in star-forming regions, such as $U_{\rm min}$ and $\gamma$ (\texttt{umin} and \texttt{gamma} in Tab.~\ref{tab:gridCIGALE}), correlate with dust emission at 24~$\mu$m, 71~$\mu$m and 160~$\mu$m \citep{Draine07}. Similarly, 60-100~$\mu$m observations can constrain the power law coefficient $\alpha$ (\citealt{Dale14}, \texttt{alpha} in Tab.~\ref{tab:gridCIGALE}). Observations of MIR-FIR integrated light from local galaxies have showed that $U_{\rm min}$ ranges between 1 and 10, while $\gamma$ yields best fit values of 0.5 - 1 \citep{Draine07b} and $\alpha$ is generally fixed to 2 \citep{Draine07}. Based on these findings we fitted the SEDs using an {\it informed (IF) grid}, where we constrain $U_{\rm min}$ to be either 1 or 10, while the allowed values for $\gamma$ are 0.5 and 1, following \citet{Draine07b}. We do not impose any additional constraints on the possible values of q$_{\rm PAH}$, as this quantity depends on the incident radiation field and local ISM conditions \citep[e.g.,][]{Chastenet19}. By inspecting the recovered dust parameter distributions (see Fig.~\ref{fig:comparison_infgrid} for M51), we noticed different results compared to the parameter grid of Tab.~\ref{tab:gridCIGALE}. For this reason, we have explored a larger grid and investigated the favored parameter space when fitting. We refer to the model grid that provides the best fit as {\it adopted (AD) grid} shown in Tab.~\ref{tab:gridCIGALE}, with less strict constraints on $U_{\rm min}$ and $\gamma$ compared to the informed one.

Since our observations are limited to the NIR spectrum, we can evaluate the effect of these parameters only on the quality of the CIGALE fits to the optical and near-infrared SEDs. In Fig.~\ref{fig:comparison_infgrid} we present a comparison between CIGALE results for eYSCI in M51 obtained with the AD grid and with an IF grid, where the only difference are the constraints on $U_{\rm min}$ and $\gamma$ (Tab.~\ref{tab:gridCIGALE}).
This figure shows that the recovered reduced $\chi^2$ is lower for eYSCI fitted with the AD grid. Additionally, when looking at the fitted age distribution, a large fraction of eYSCI fitted with the IF grid settles at the edge of the parameter space, returning either young ages or old ones. Moreover, as a side check, we note that our results for the AD grid remain consistent even when we additionally include $U_{\rm min} = 10$ and $\gamma = 1$.

Hence, we conclude that our AD grid gives better results in terms of quality of the fits. We must stress that since we cannot constrain the dust parameters, there is no physical reason associated with the choice of the dust grid behind this consideration (see Sec.~\ref{sec:dustDisc} for a further discussion). However, this unexpected behavior calls for a careful analysis of the eYSC SEDs, trying to unveil the reason behind this result (see also Linden et al.~(in prep.), for a more detailed exploration for eYSCs in NGC~628).


\subsection{CIGALE grid for optical catalogs}\label{sec:optCIGgrid}

Due to the different nature of optically selected YSCs (see Sec.~\ref{sec:opticalCat}), the CIGALE adopted grid for these specific objects requires small modifications from the eYSCs grid presented in Tab.~\ref{tab:gridCIGALE}. Since these clusters are identified in $V$ band, there are no prior constraints on their age and therefore we allow the \texttt{age} parameter vary from 1 Myr to 14 Gyr. Additionally, optically selected clusters present lower value of E(B-V) (e.g. \citealt{Whitmore20}, \citealt{Knutas25}). For this reason, we limit the \texttt{E\_BV\_lines} parameter grid to 1.5. In the specific case of NGC~4449, the metallicity grid was extended to include 0.004 in addition to 0.008. For the optical YSC catalog in NGC~4449, Appendix~\ref{app:opt4449} presents a comparison between the ages we derived using CIGALE in this work and those recovered by \cite{Whitmore20} from the LEGUS survey for the same objects, showing good agreement when filters that include hydrogen emission lines are used in the fit.

\subsection{Fitting results} \label{sec:CIGres}

In this section, we present an overview of the results obtained with CIGALE for eYSCI in our sample of galaxies. Final photometric catalogues and CIGALE SED fit outputs will be released at \url{https://feast-survey.github.io}. We listed the number of eYSCI for each galaxy in Tab.~\ref{tab:targets}. 
\begin{figure*}[htb]
    \centering
    \includegraphics[width = 1.0\textwidth]{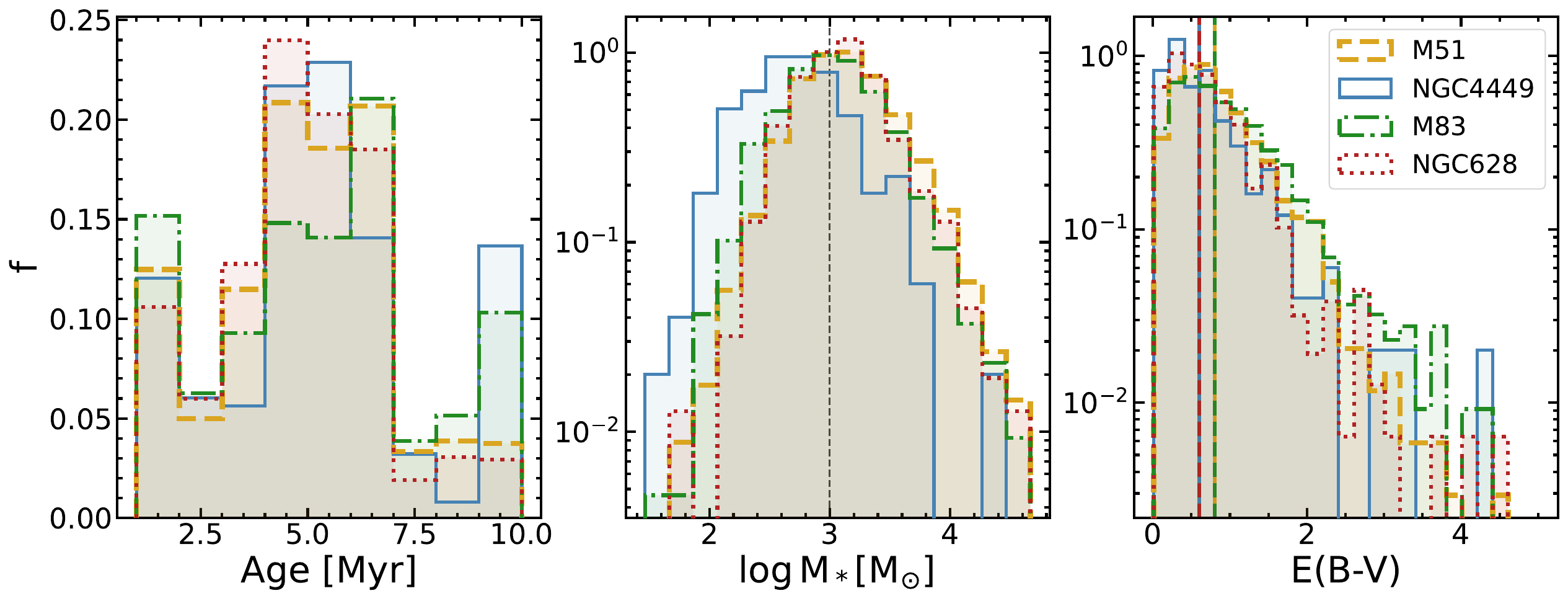}
    \caption{Normalized distributions of age, stellar mass, and E(B-V) for eYSCs in our sample of galaxies, obtained with CIGALE. We use different colors and line styles to distinguish the distributions of each galaxy. In the middle panel, we drew a dashed line at M$_*=1000$ M$_{\odot}$ to guide in the plot visualization. In the right panel, we added median values of the E(B-V) distributions as vertical lines. }
    \label{fig:summary_physProp}
\end{figure*}
Fig.~\ref{fig:summary_physProp} illustrates the results of the best recovered values for age, stellar mass and attenuation E(B-V). Histograms are color-coded by different galaxies. These plots show that eYSCs in our sample of galaxies share similar behavior in the distributions. The age distributions exhibits a steep drop after 6 Myr, due to the fact that older clusters produce a considerable smaller fraction of ionizing photons than the younger counterpart \citep{Leitherer99,Calzetti13}. The mass distributions show similar trends between each other. NGC~4449 presents overall lower masses. We observe a similar scenario in the E(B-V) distributions. In this case, M51 and M83 show higher E(B-V), with median values of about 0.8 mag., while NGC~628 and NGC~4449 show slightly lower values with medians of 0.6 mag for both distributions, consistent with what found by \cite{Gregg24} and Gregg et al.~(subm.) for these galaxies using the H$\alpha$/Pa$\alpha$ ratio. 

\begin{figure*}[htb]
    \centering
    \includegraphics[width = 0.496\textwidth]{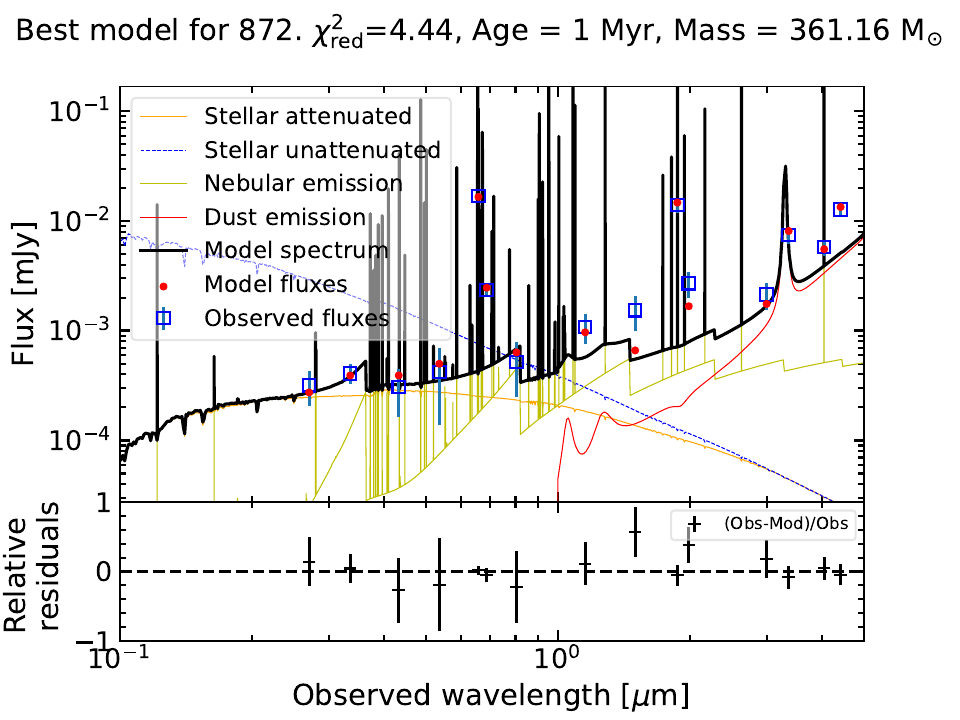}
    \hfill
    \includegraphics[width = 0.496\textwidth]{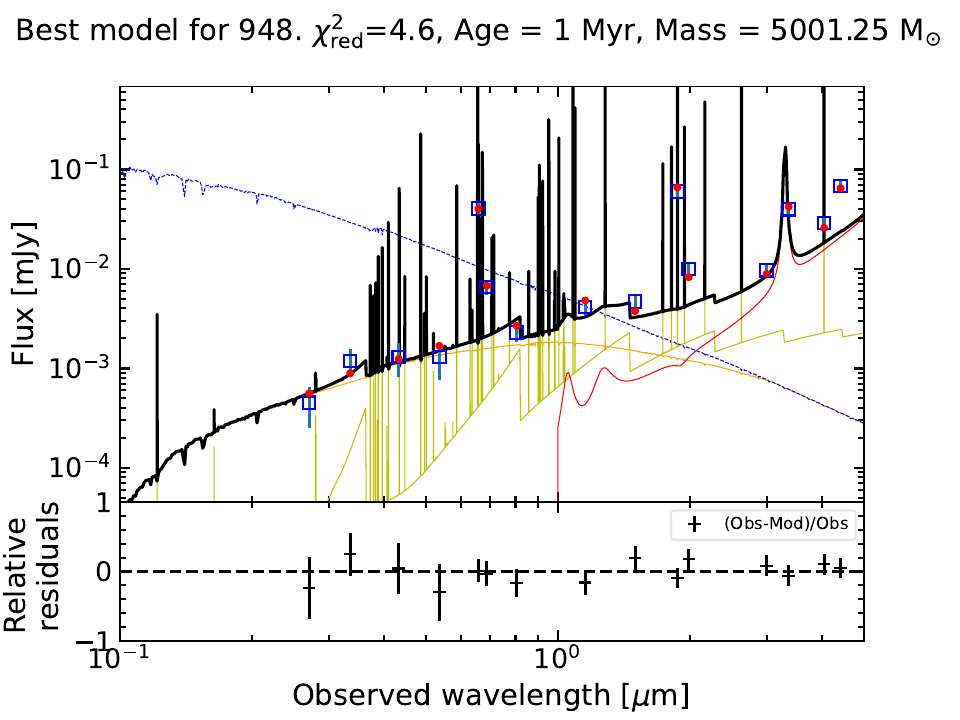}
    \\[\smallskipamount]
    \includegraphics[width = 0.496\textwidth]{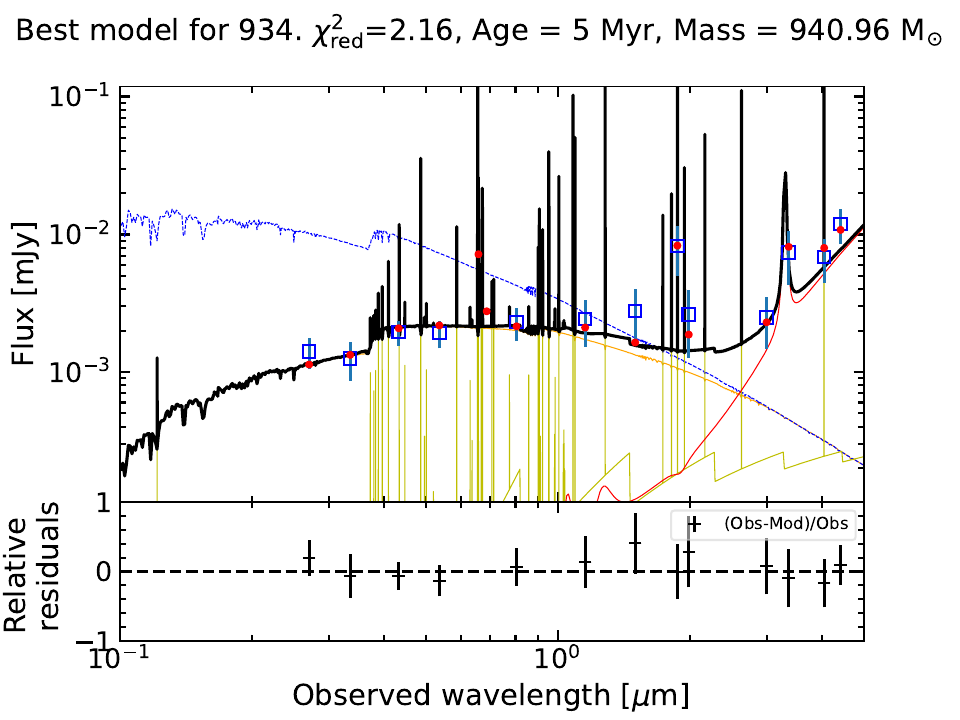}
    \hfill
    \includegraphics[width = 0.496\textwidth]{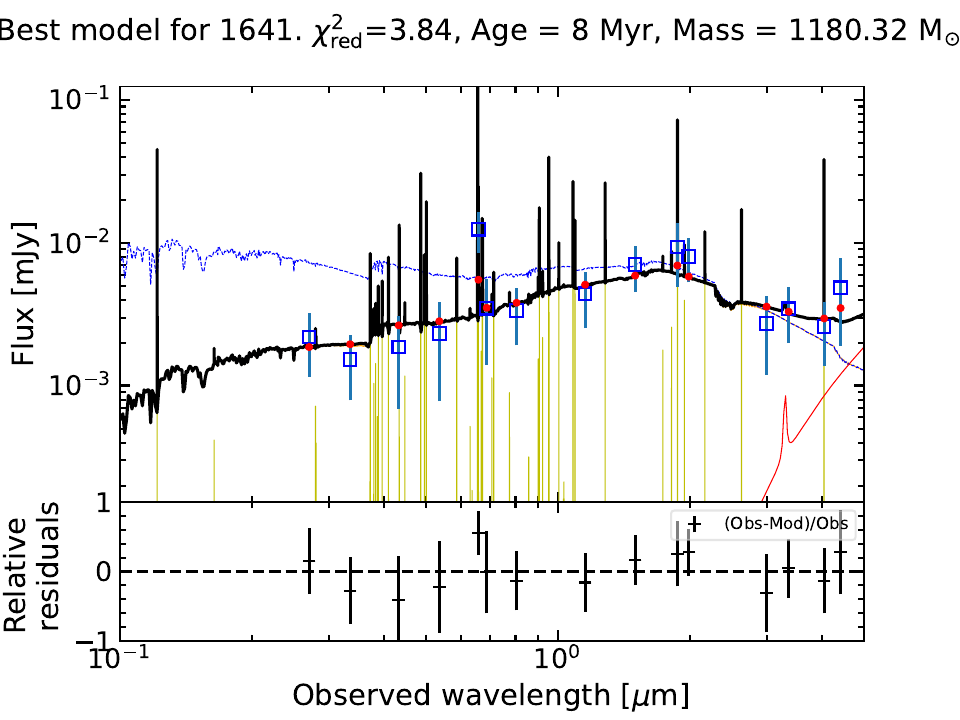}

    \caption{Observed SEDs, best fit models and recovered best fluxes for four representative eYSCs in M51. For each panel, the title of the plot describes the fitted properties of the specific eYSC (represented by a unique ID): $\chi^2_{\rm reduced}$, Age and stellar mass. The observed fluxes in HST and JWST filters are shown with blue squares with error bars, while we show the CIGALE best fitted fluxes convolved to each filter throughput with red filled circles. The black line represents the best model spectrum, while the remaining colored lines correspond to best fitted models for the attenuated and unattenuated stellar component (orange and blue spectra, respectively), nebular (green spectrum) and dust emission (red spectrum). For each filter, we show the relative residual fluxes (obs - best)/obs in the lower portion of the panel.}
    \label{fig:SEDs_ex}
\end{figure*}

To better understand the meaning of these results, we visually inspected a set of representative SEDs from the CIGALE output, where models are compared with observations. In Fig.~\ref{fig:SEDs_ex} we present four representative examples of eYSCI SEDs in M51. In this figure, the observed fluxes in each filter (blue unfilled squares) are compared to the corresponding best flux values recovered by CIGALE (red filled circles). Underneath those points, we show the best model grid returned by the code. We show the residuals of the fit in the bottom panel as (obs - best)/obs, where obs is the observed flux and best is the best fitted value, for each filter. The title of each plot describes the fitted properties of the given eYSCI in term of id, best reduced $\chi^2$ value, best age and best stellar mass. 

Already from these representative cases, we note that regardless of the reduced $\chi^2$ values, the CIGALE modeling is not able to fully reproduce the SED of the clusters. We find that the quality of the fit depends on the recovered age and mass of the eYSC. In the top left panel, we present the best fitted model for an eYSC with a fitted age of 1 Myr and M$_*$ of 361.16 M$_{\odot}$. Even though the recovered $\chi^2_{\rm red}$ suggesting a good quality of the fit, it becomes evident that some NIR broad bands appear underestimated in the CIGALE fit. In this case, the observed fluxes are systematically higher than the best fit values, even when accounting for the uncertainties. This issue is particularly evident for the F150W filter and the offset between the best fit values and the observations seems to decrease in amplitude (or even disappear) going towards either higher masses (top right panel of Fig.~\ref{fig:SEDs_ex}) or older ages (bottom panels of Fig.~\ref{fig:SEDs_ex}). This discrepancy is what we will refer in the next section as the NIR excess.

\section{NIR excess} \label{sec:NIRexcess}

The CIGALE modelling presented in the previous section reveals the presence of an observed NIR flux excess that the best fit models cannot recover. In this section, we further quantify this excess over the entire sample of eYSCI. 
\begin{figure*}[htb]
    \centering
    \includegraphics[width = 0.85\textwidth]{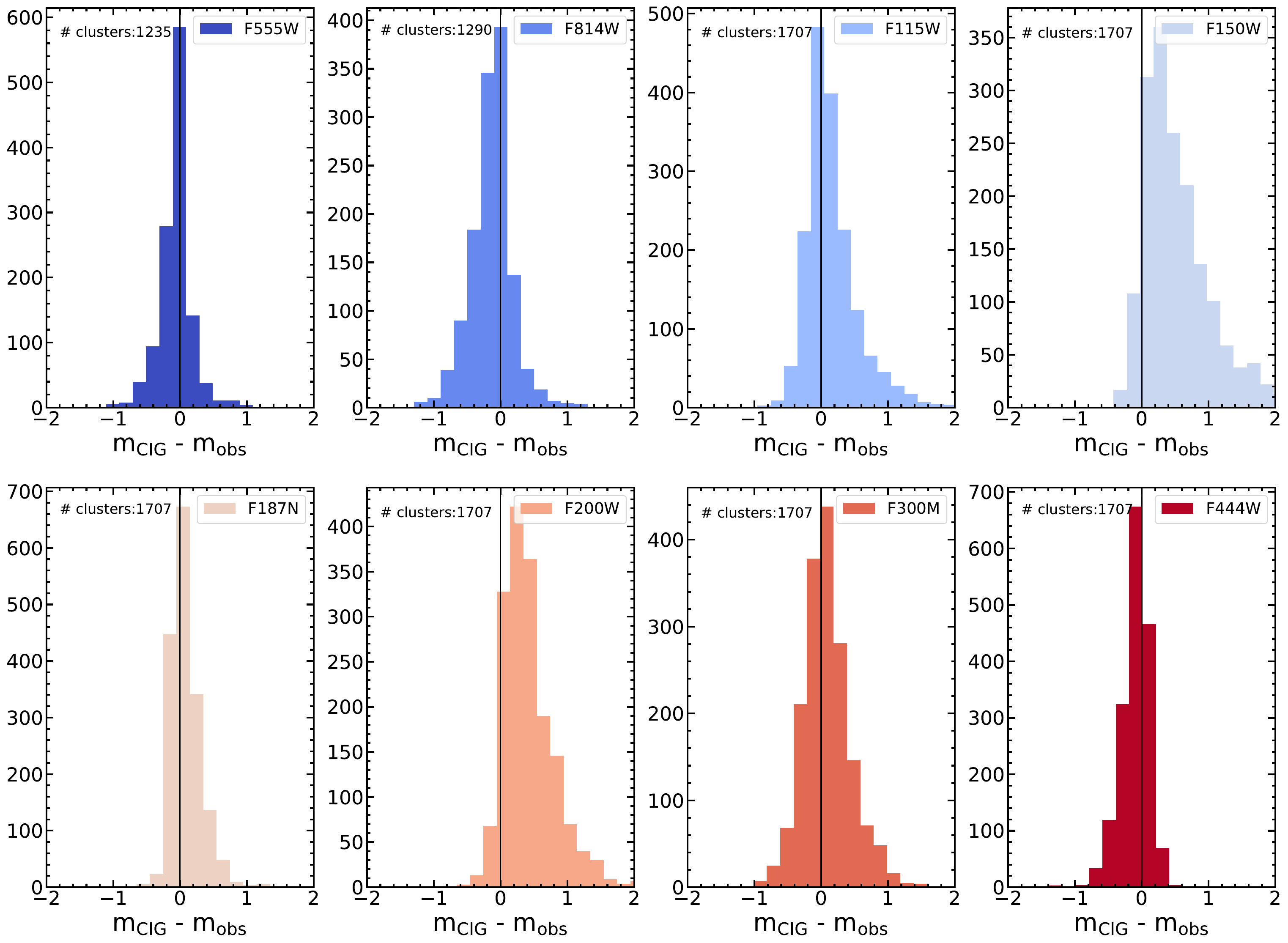}
    \caption{Distributions of the residuals m$_{\rm CIG}-$m$_{\rm obs}$ in eight selected filters (F555W, F814W, F115W, F150W, F187N, F200W, F300M, F444W) for the sample of eYSCI in M51. For each filter, m$_{\rm CIG}-$m$_{\rm obs}$ represents the difference in magnitude between the best fitted flux from the CIGALE fit and the observed flux, respectively. For each panel, we show the total number of eYSCI, which varies in HST filters due to the additional S/N requirements described in the text. To guide the reader's eye, we plot a black vertical line at $m_{\rm CIG} - m_{\rm obs} = 0$.    }
    \label{fig:summary_residuals}
\end{figure*}

We calculate the residual difference between the magnitude returned by CIGALE, m$_{\rm CIG}$, and the observed magnitude, m$_{\rm obs}$. For this analysis we limit ourselves to eYSCI that are well detected in all the NIRCam bands (see Sec.~\ref{sec:eyscPresentation}). If a specific HST-UV/optical magnitude is included in the residual analysis, we require that the eYSCI have been detected with a S/N $>3$ in that filter. 

We use the M51 cluster population as a reference, but we apply the same analysis to all FEAST galaxies introduced above. In Fig.~\ref{fig:summary_residuals}, we plot the distributions of the residuals, i.e., the difference m$_{\rm CIG}$ $-$ m$_{\rm obs}$ for the following filters: HST/F555W, F814W and JWST-NIRCam/F115W, F150W, F187N, F200W, F300M, F444W. We note that the number of eYSCI in the HST bands is lower than the number in the four NIRCam bands, due to the S/N requirement we applied for the single HST bands. For each filter, a distribution centered on $\Delta m\simeq 0$ would indicate a good agreement between models computed by CIGALE and observations. On the other hand, a prominent tail in the distributions at $\Delta m>0$ would indicate that the observed flux in that filter is brighter that the one predicted by the best-fitted model, i.e. the latter model underestimates the recovered flux.

The $\Delta m$ distributions in Fig.~\ref{fig:summary_residuals} show some interesting trends. In general, we see that the fluxes in the HST/F555W, NIRCam/F187N, F300M, F444W, are centred around zero (see the relative numbers of objects in the peak ($\Delta m\pm0.2$ mag) with respect to the wings). On the other hand, the distributions in NIRCam/F150W and F200W show indeed a prominent tail going towards positive values of $\Delta m$, meaning that the best-fitted models cannot properly reproduce the observed fluxes in a large fraction of systems. The HST/F814W residuals appear to suffer from the opposite behavior, i.e. the best solution that attempts to fit the NIR excess would under-perform in the bluer band. These widespread trends, along with the "unconstrained" grid of dust parameters discussed above, suggest an underlying severe problem for CIGALE to reproduce the observed SEDs of the eYSCs.

\begin{figure*}[htb]
    \centering
    \includegraphics[width = 0.496\textwidth]{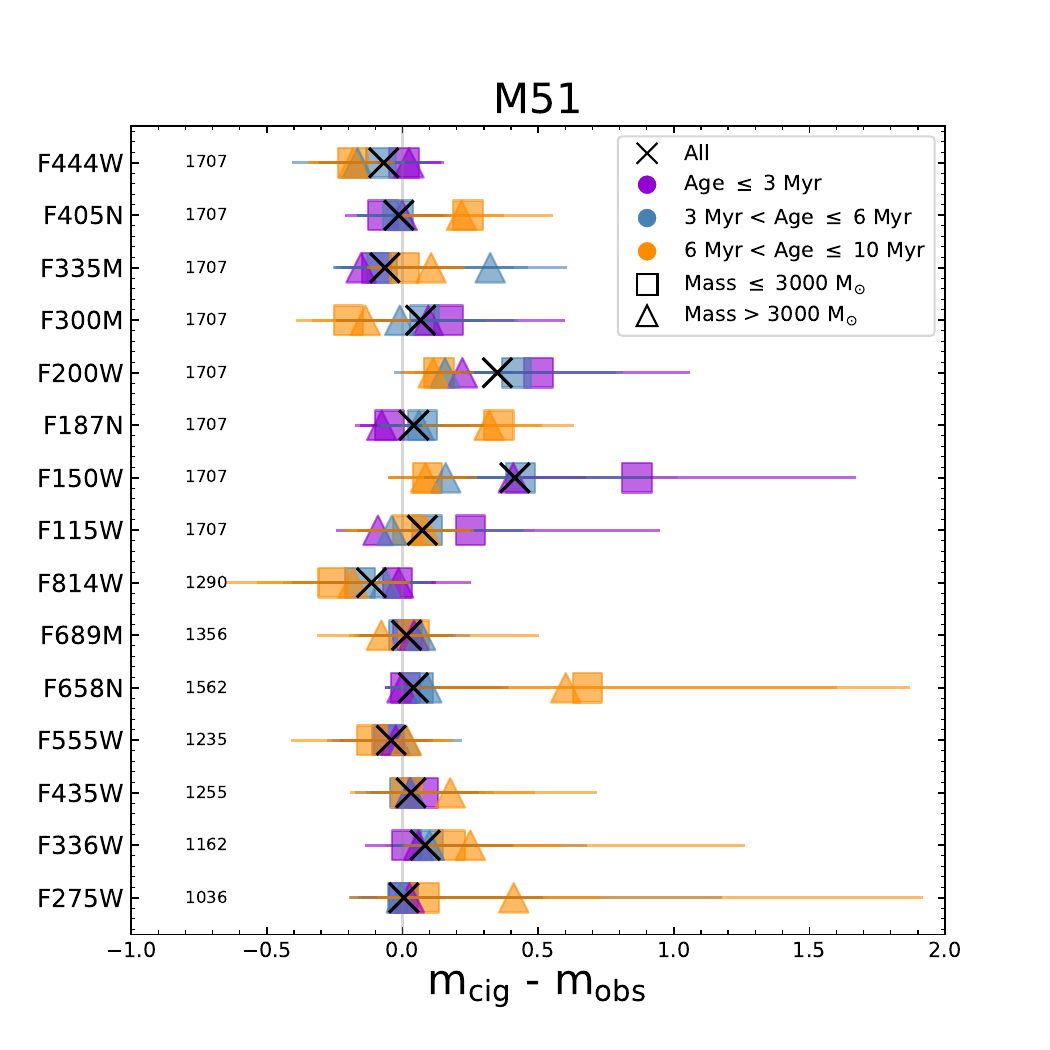}
    \hfill
    \includegraphics[width = 0.496\textwidth]{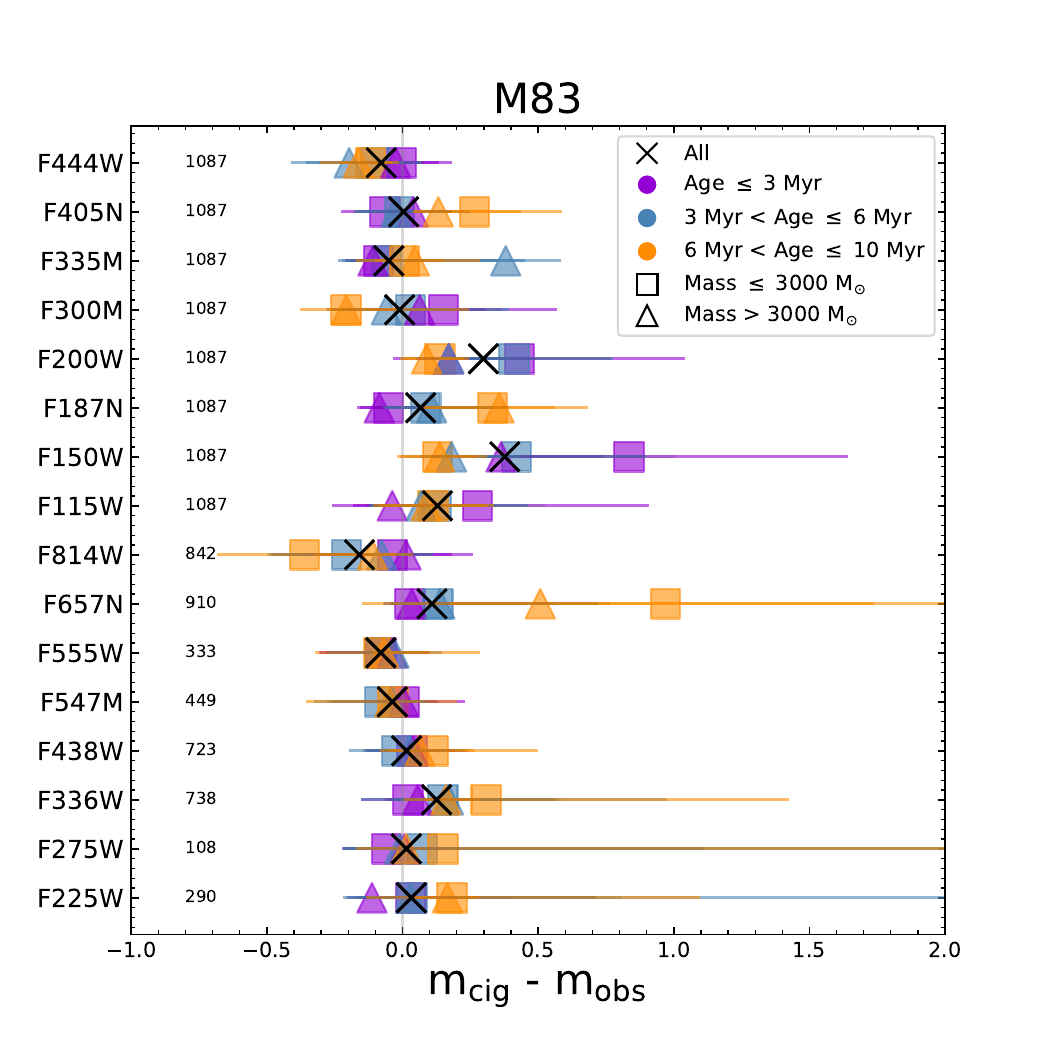}
    \\[\smallskipamount]
    \includegraphics[width = 0.496\textwidth]{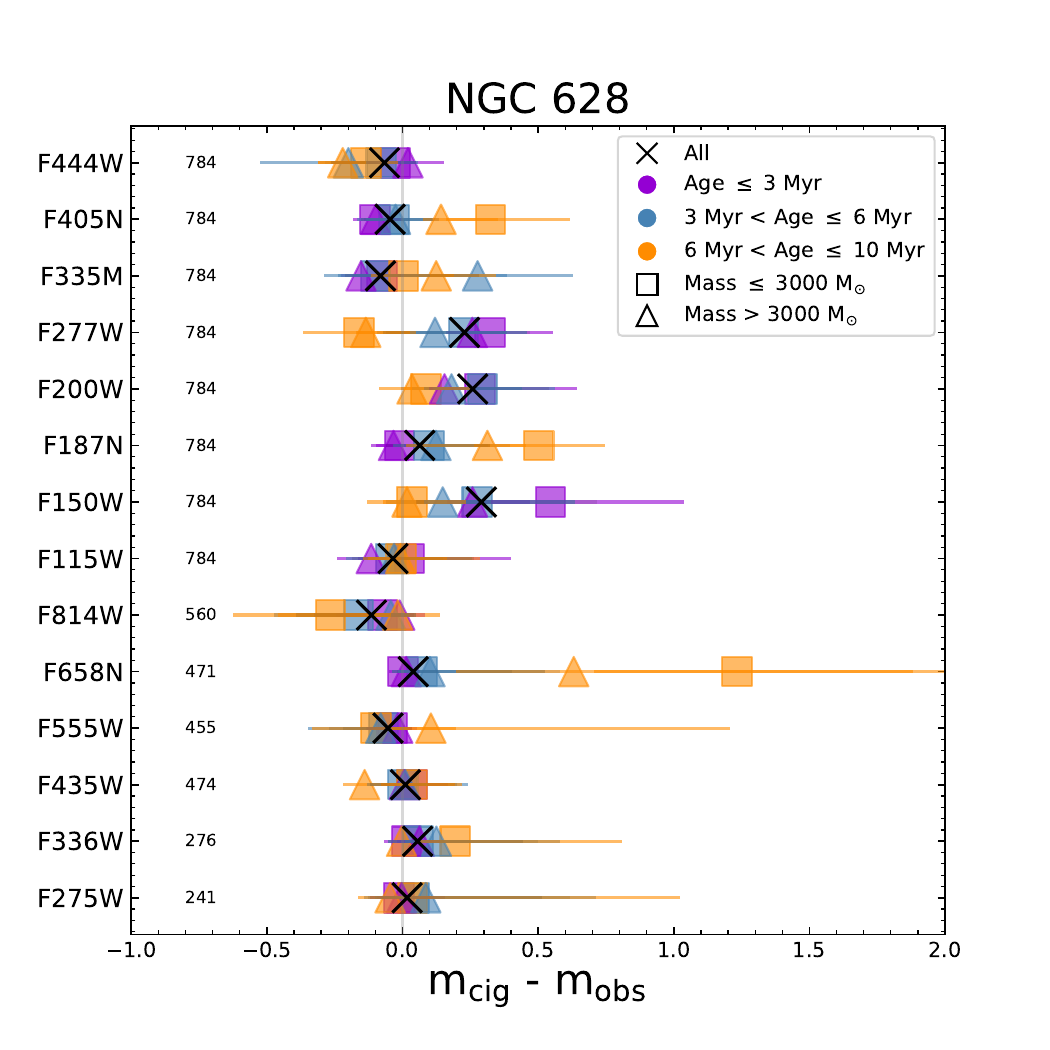}
    \hfill
    \includegraphics[width = 0.496\textwidth]{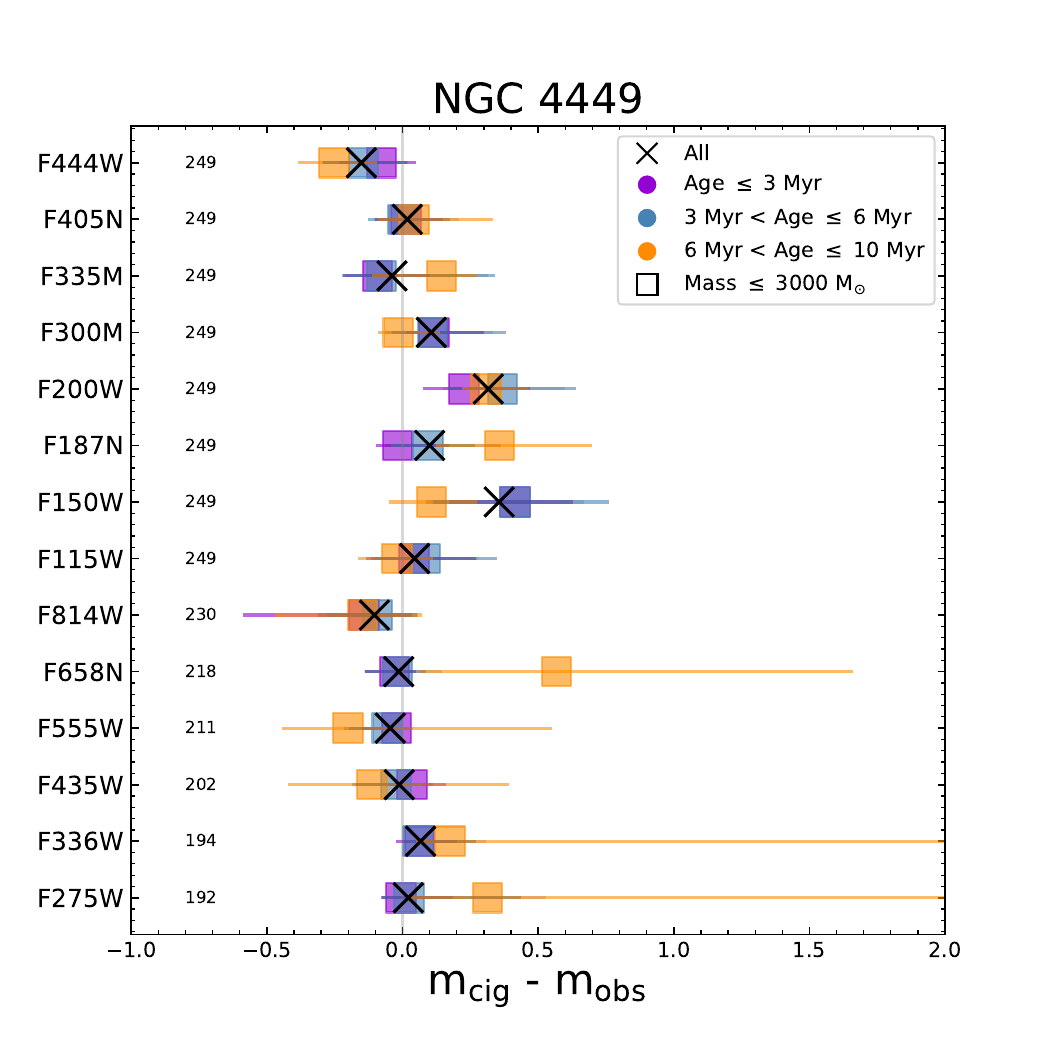}
    \caption{Median values of the residual distributions m$_{\rm CIG}-$m$_{\rm obs}$ (see Fig.~\ref{fig:summary_residuals}) for each available filter in eYSCI across the four galaxies studied in this work. For each panel, black crosses represent the corresponding median value of m$_{\rm CIG}-$m$_{\rm obs}$ in that specific filter, for the entire population of eYSCI (where the total number of clusters is specified in the plot for each filter). HST filters present smaller numbers of clusters due to the additional S/N criteria we applied in this analysis. Different colors represent sub-samples of the population of eYSCI binned by the best fitted age recovered with CIGALE. Additionally, we use two different symbols (square and triangle) to distinguish between two bins of masses ($\leq 3000$ M$_{\odot}$, $> 3000$ M$_{\odot}$). For each data point, error bars correspond to the 16th-84th percentile of the residual distribution. Every galaxy in our sample exhibits an excess in the NIR, particularly more offset can be seen for the low mass and young ($\leq$ 3 Myr) bins.}
    \label{fig:NIRexcess}
\end{figure*}

To better investigate this result, we unfold the residuals in each available filter of each galaxy by using the recovered ages and masses of the eYSCs. We present the median values and 16th-84th percentiles of the $\Delta m$ in Fig.~\ref{fig:NIRexcess}. We divide our samples of eYSCI in three age bins (1-3 Myr in purple, 3-6 Myr in blue, 6-10 Myr in orange) and two mass bins (M$_* \leq 3000$ M$_{\odot}$ as squares and M$_* > 3000$ M$_{\odot}$ as triangles). The three age bins represent three different evolutionary stages of eYSCI, while the two mass bins roughly distinguish between cluster that possibly fully sample the IMF (higher mass bin) and clusters that might be subjected to stochastic sampling effects \citep{Cervino04}. In general, all the sub-samples defined in this analysis are statistically significant, with the number of eYSCI in each bin ranging from tens to hundreds. For the three spirals, $\sim$ 80\% of the eYSCI have fitted masses $\leq 3000$ M$_{\odot}$, and 40 \% have masses below $\sim 1000$ M$_{\odot}$. In the case of NGC~4449, we omitted the high mass bin, as this galaxy shows only 16 eYSCI with fitted mass higher than 3000 M$_{\odot}$. As a guidance, we also include the median values of the whole sample as black crosses.

The medians of the full populations (black crosses), as already demonstrated in Fig.~\ref{fig:summary_residuals}, show an observed NIR excess in the broad-band filters F150W and F200W of all the targets, and in F277W of NGC~628 (the only galaxy for which this filter is available at the time of this work). The F300M, which replaced the F277W, does not seems to be significantly affected. As we look at the residuals per cluster age and mass bins, some interesting trends are evident. Clusters fitted with ages younger than 6 Myr and mass less than 3000 M$_{\odot}$ (purple and blue filled squares) appear to be the ones driving most of the recovered positive residuals in the F150W and F200W (and F277W). In particular, the excess is highest in F150W for the low mass clusters younger than 3 Myr (purple square). In all galaxies, the eYSCI fitted with ages older than 6 Myr (orange symbols) are the ones with the largest residuals in all the narrow bands. Particularly telling for this group is the inability of CIGALE to provide a best fit model that reproduces the excess in the narrow band filters transmitting H$\alpha$ (HST/F657N and F658N), Pa$\alpha$ (NRCam/F187N), Br$\alpha$ (NIRCam/F405N). We conclude that the recovered physical properties for this ("older") group are not correct and due to the presence of strong ionized hydrogen line emissions these clusters are likely younger. We additionally checked for trends between the observed NIR excess in the F150W filter and the attenuation E(B-V) of the clusters, finding no significant correlation between them (Pearson correlation test: r = -0.075, p = 0.002).

Another important consideration to make is that while we observe similar trends in all groups and average populations of all the galaxies of the sample, the amplitude of this model-observed difference appears to vary as a function of the host galaxy. In particular, the magnitude in the NIR excess in the young and low mass bin in NGC~4449 is lower than the one observed in the spiral galaxies, and among the spirals, the residuals in the eYSCI population of NGC~628 are slightly smaller.


\section{Discussion} \label{sec:discussion}

The recovered NIR excess presented in the previous section brings more uncertainties and makes the interpretation of the physical parameters derived with CIGALE (e.g., Fig.~\ref{fig:summary_physProp}) more challenging. As noticed in the previous section, the inability of the models to reproduce the observed eYSC SEDs results in errors in the recovered cluster ages (up to a factor of $\sim 2$, due to the presence of hydrogen emission lines) and may lead to incorrect conclusions about star formation timescales.
Here we first test whether the parameter choice to perform photometry might be at the root of the problem, and we discuss/test other more physical reasons. 

\subsection{The effect of extraction aperture on the recovered cluster SEDs}
It is well known in the literature that the use of large apertures when studying star clusters leads to severe contamination, especially in the NIR bands \citep{Bastian14}. 
As star cluster sizes range on average between 2 and 4 pc \citep{Krumholz19}, using apertures of the order of 5-10 times the cluster size results in a flux excess from contaminants which are not necessarily associated to the cluster emission. This effect appears to be stronger in the NIR wavelengths where red super giant stars (RSGs) can contaminate the photometry \citep{Bastian14}.

The apertures in our study, however, are much smaller and provide cluster photometry at a physical scale of 3-6 pc and therefore we are confident the photometry presented in this work does not present significant contamination from external sources. Nonetheless, we further verified the aperture effect and tested apertures of 4, 5, and 6 pixels to check whether the excess strongly increases for larger apertures. In Appendix~\ref{app:aperture}, we show the representative results of this analysis in M83, one of the closest galaxies in our sample with a high number of detected eYSCs. We select the smallest radius to be 4 pixels, since this aperture size is still larger or comparable to the full width at half maximum of the reddest filter F444W ($\sim 3.6$ pixels, \citealt{Rigby23}). This test confirms that the excess residuals get stronger for apertures of 6 pixels ($\sim 5.5$ pc). For the closer galaxies (M83 and NGC~4449), using either 4 or 5 pixels does not imply a significant change, while for the more distant ones (NGC~628 and M51) the excess residual get stronger at 5 pixels. For these reasons, we adopted photometric radii of 4 pixels for the more distant galaxies in our sample, while we used 5 pixels for the closest ones. In conclusion, increasing the aperture results in a more pronounced NIR excess, as already pointed out in the literature, but the excess is still significant when small apertures are adopted.

\subsection{Testing dust models on the SED fit analyses}
\label{sec:dustDisc}
Another source of uncertainty might come from the inability of our data to inform CIGALE on the dust properties. While JWST offers MIR coverage up to 20 $\mu$m, the reduced spatial resolution implies an increase in the size of the adopted photometric aperture, and therefore an increase in the NIR excess and contamination discussed in the previous Section.
As already described in Sec.~\ref{sec:dustgrid}, the resulting dust parameter grid we use is the one that gives the lowest values of $\chi^2$ in the CIGALE outputs, but the ranges are quite off from what we would have expected from MIR-FIR constraints (e.g., $U_{\rm min}$ and $\gamma$, see Sec.~\ref{sec:dustgrid}). It is perhaps likely that also this behavior in the models is driven by the presence of the NIR excess: CIGALE "prefers" unrealistic dust models in the attempt to compensate for a signal that cannot be fitted only with standard single stellar population models and nebular emission. This might be expected when a model is missing one or more components \citep[e.g.,][]{Peck80}. To inspect this possibility, we performed a new SED fitting analysis for the eYSCI population in M51, this time with the exclusion of the dust grid from the CIGALE parameter space. Moreover, in the reasonable approximation that PAH and dust emission start to dominate the spectra of star clusters after 2-3 $\mu$m, we fitted the eYSCI up to the NIRCam F200W filter. With this method, our SED best fit values are exclusively constrained by stellar and nebular emission, at wavelengths below $\sim$2 $\mu$m.

\begin{figure}[htb]
    \centering
    \includegraphics[width = 0.496\textwidth]{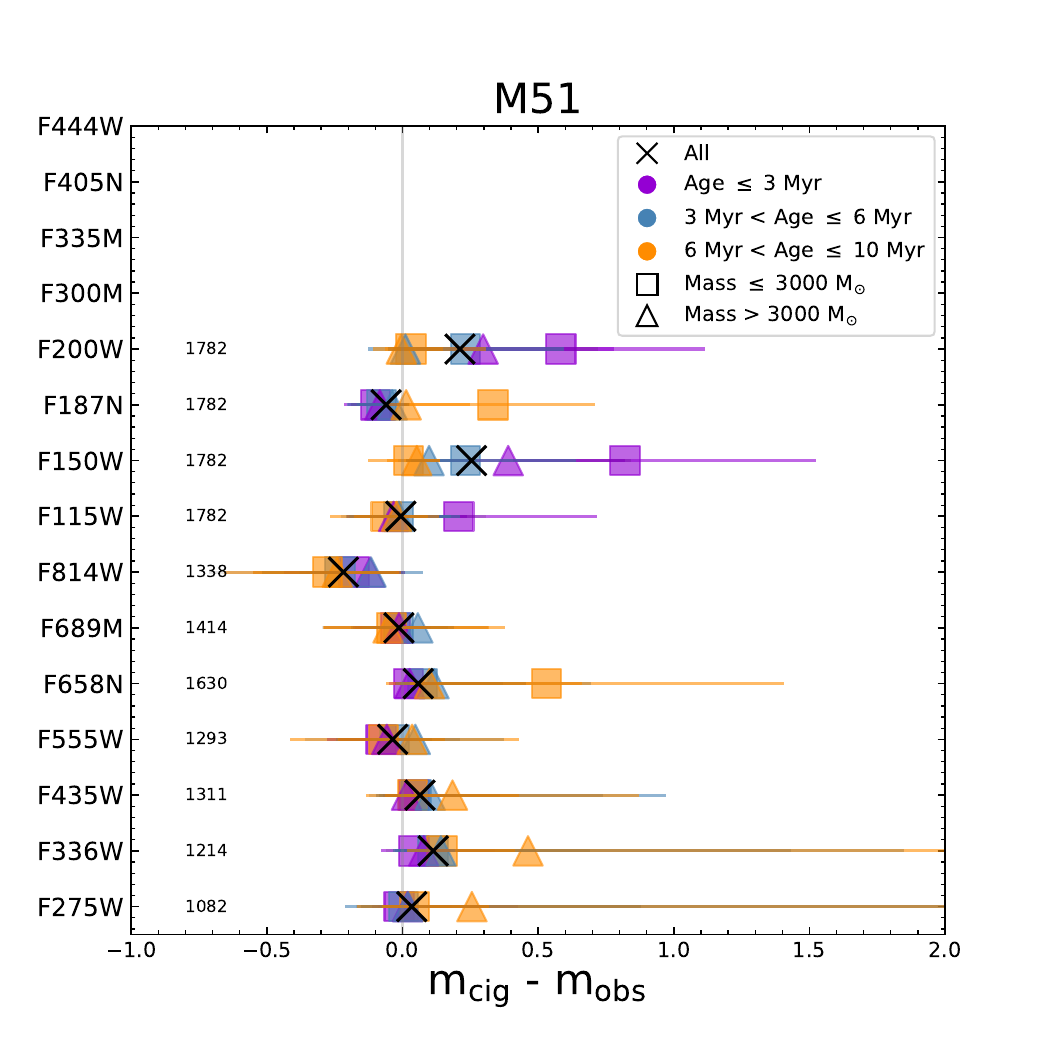}
    \caption{Median values of the residual distributions m$_{\rm CIG}-$m$_{\rm obs}$ for eYSCI in M51, as shown in the top left panel of Fig.~\ref{fig:NIRexcess}. In this case, the CIGALE SED fit is performed up to the F200W filter and ignoring the contribution from the dust models.}
    \label{fig:up2F200W}
\end{figure}

In Fig.~\ref{fig:up2F200W}, we show the residual analysis in M51 for this new SED fitting approach. 
We note that the number of eYSCI recovered in the NIRCam filters is slightly higher than the complete fit. This is due to the fact that we are considering a smaller number of NIRCam filters when applying the S/N cut defined in Sec.~\ref{sec:eyscPresentation}. The NIR excess is still evident and exhibits similar trends discussed in Fig.~\ref{fig:NIRexcess}, regardless of whether dust models are included in the fitting procedure This result excludes the unconstrained choice of the dust parameter grid as a driver for the observed NIR excess.

\subsection{The NIR excess in optical YSCs}
Taken at face value, the change in the strength of the excess discussed in Sec.~\ref{sec:NIRexcess} might suggest a relation with age. We therefore investigate the residuals in the optically selected YSC population with fitted age younger than 10 Myr in M51 (with the entire parameter grid and set of filters). As described in Sec.~\ref{sec:opticalCat}, YSCs are identified as bright source of emission in optical bands. Since they are more evolved and less dusty, we expect them to show a limited contribution from emission in the NIR. Clusters where their SED is dominated by stellar and nebular emission are ideal for understanding how dust models are influencing the results. For the scope of this analysis, and to avoid adding further uncertainties, we select only YSCs detected in all NIRCam bands with S/N $> 3$, analogous to what was presented for the eYSCs. We also remind the reader that to be classified as YSCs (see Sec.~\ref{sec:opticalCat}), they must be detected in at least 4 HST bands as well.

\begin{figure}[htb]
    \centering
    \includegraphics[width = 0.496\textwidth]{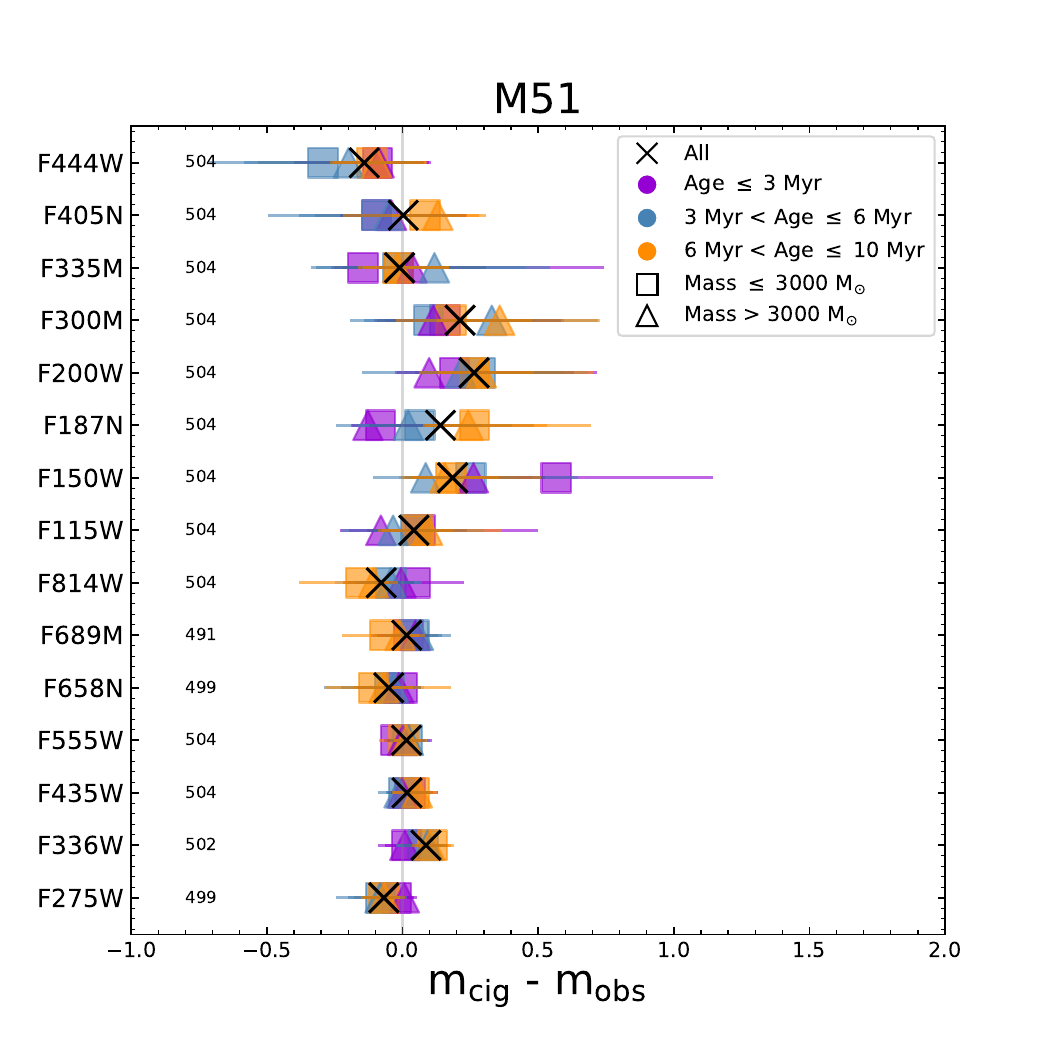}
    \caption{Median values of the residual distributions m$_{\rm CIG}-$m$_{\rm obs}$ for the optically selected YSC population in M51 younger than 10 Myr. The layout of the figure is as presented in Fig.~\ref{fig:NIRexcess}.}
    \label{fig:res_OPT}
\end{figure}
We present the results of the residual analysis for this population in Fig.~\ref{fig:res_OPT}. 
It is evident that the NIR excess is less pronounced in optically selected YSCs in M51, with median residuals going from about 0.5 to 0.2 mag. Moreover, we see that differently from the eYSCI (see Sec.~\ref{sec:NIRexcess}), the hydrogen recombination lines are well fitted. In Appendix~\ref{app:optapp}, we confirm these results also for the other three galaxies in the sample.

\subsection{The impact of stochastically sampling the IMF} \label{sec:slug}
The discrepancy between models and observations observed in this work is not uniform across clusters. As shown in Fig.~\ref{fig:NIRexcess}, clusters in bins characterized by younger ages and lower masses are more likely to show a NIR excess, while more massive clusters show smaller residuals overall. 
One of the possible reasons behind this behavior may be linked to IMF sampling issues. For stellar masses below $10^4$~M$_{\odot}$ \citep{Cervino04}, stochasticity in the IMF sampling becomes important and it cannot be ignored at masses of a few hundreds of M$_{\odot}$ \citep{Calzetti10}. 
For cluster mass ranges where stochasticity becomes important, the number of massive stars, dominating the integrated light and therefore the derived ages and masses, might vary strongly from cluster to cluster with the same mass. For instance, clusters with total stellar masses ranging from a few hundred M$_{\odot}$ to 1000 M$_{\odot}$ contain between a few and a few tens of massive stars (M $> 8$ M$_\odot$), respectively \citep[see Fig.1 in][]{Stanway23}. Furthermore, while the optical wavelengths are typically dominated by the massive stars SED, the NIR emission has also an important contribution from pre-main-sequence (PMS) stars.
PMS stars are in an evolutionary stage where the star is still contracting towards the main sequence, making them bright sources in the 1-2.5 $\mu$m NIR wavelengths \citep{Heyer90,Kenyon95,Eiroa01}. In stellar clusters where stochastic effects are non-negligible (stellar masses $\lesssim 10^4$~M$_\odot$), the variation in the number of massive stars can significantly impact the integrated NIR light. In such cases, the emission from PMS stars may contribute substantially to the observed NIR flux. However, CIGALE employs a deterministic approach in which the IMF is fully sampled, normalized to 1 M$_{\odot}$, and then rescaled to reproduce the observed SED \citep{Boquien19}.
Moreover, stellar spectra models from \cite{Bruzual03} do not account for PMS stars, which means CIGALE cannot take into account their contribution.

Investigating the impact of a stochastic approach on estimating the NIR colors of eYSCs requires a SED model that includes stochastic sampling of the IMF of star clusters. To reach this goal, we generated star clusters libraries using the stellar population synthesis code \texttt{slug} \citep[Stochastically Light Up Galaxies,][]{daSilva12}. Given a fixed IMF and a set of parameters that regulate stellar and nebular contributions, \texttt{slug} simulates the spectrum and the photometry of a single star cluster, where its stellar population is stochastically sampled from a probability distribution function. For star clusters, \texttt{slug} simulates a single burst of star formation with mass either fixed or drawn from a cluster mass function. Consequently, the code follows the evolution of the cluster during time and computes stellar and nebular emission \citep{Krumholz15}. We note that \texttt{slug} does not include dust models, but allows us to test the NIR colors where we see the excess.

For a fixed star cluster mass and for each age bin, the output of the simulation consists of photometric data from a given set of filters, in ABmag. 
In terms of this paper's scope, we run 10$^4$ star cluster models for a set of five fixed values of mass: 500 M$_{\odot}$, 1000 M$_{\odot}$, 3000 M$_{\odot}$, 5000 M$_{\odot}$ and 10000 M$_{\odot}$. The choice of parameters we adopted for \texttt{slug} simulations is based on maintaining the highest possible level of consistency with the grid adopted in the CIGALE analysis.
For each model, the simulation starts at 1 Myr and it ends at 10 Myr, with a time step of 1 Myr. We used a \cite{Chabrier03} IMF, where the minimum mass for the stochastic treatment is set at 0.08 M$_{\odot}$. Stellar tracks, including emission from PMS stars and rotation, are from \cite{Dotter16}, and models for computing stellar atmospheres are based on starburst99 \citep{Leitherer99}. Furthermore, \texttt{slug} uses CLOUDY based templates \citep{Ferland13} to account for nebular emission. We fixed the ionization parameter $\log U$ (see Tab.~\ref{tab:gridCIGALE}) to -2.5 and the fraction of ionized photons contributing to nebular emission, $\phi$, to 0.73. We do not apply any extinction correction to the output photometry, as this quantity varies across different clusters. 
\begin{figure*}[htb]
    \centering
    \includegraphics[width = 0.496\textwidth]{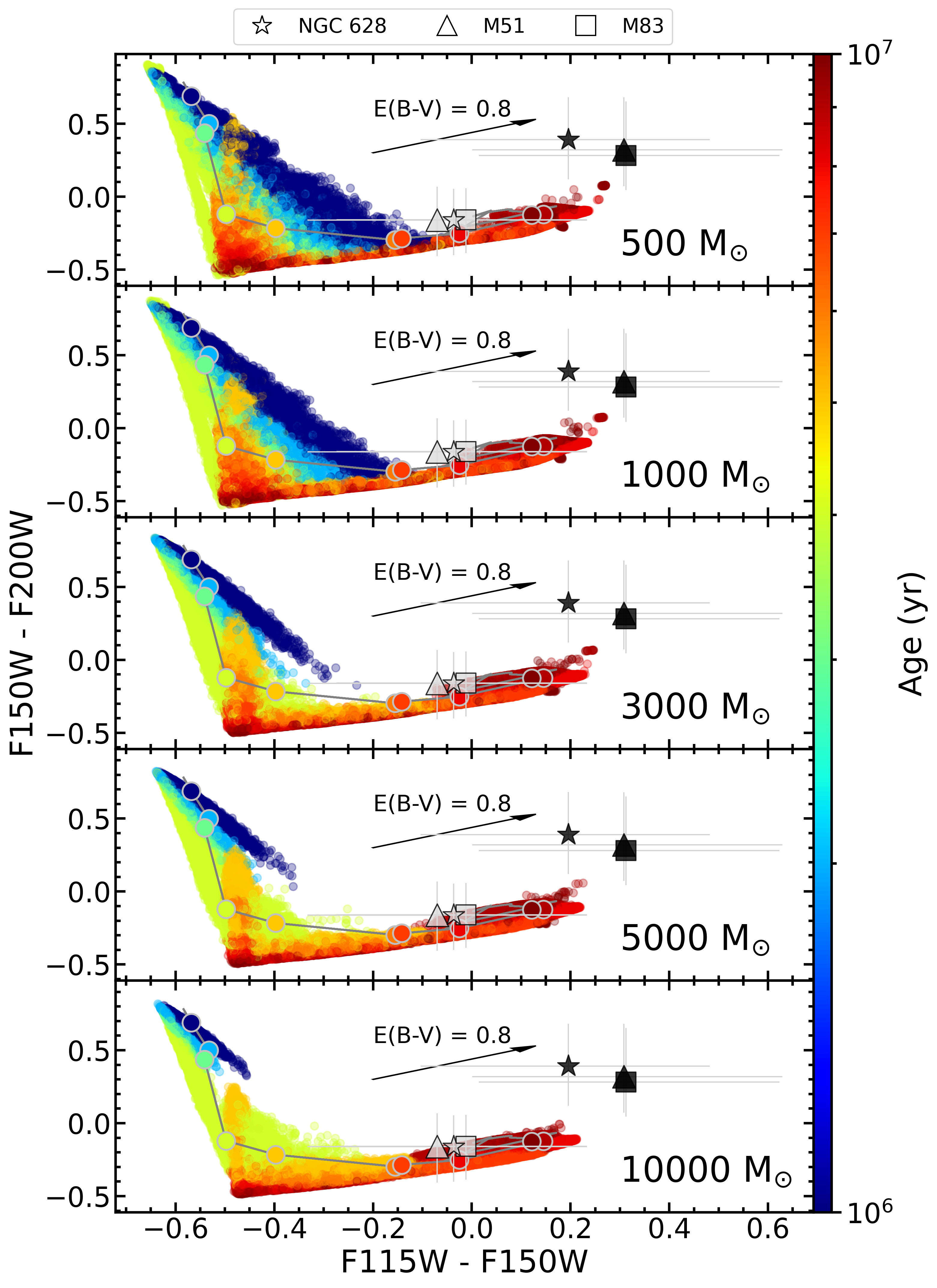}
    \hfill
    \includegraphics[width = 0.496\textwidth]{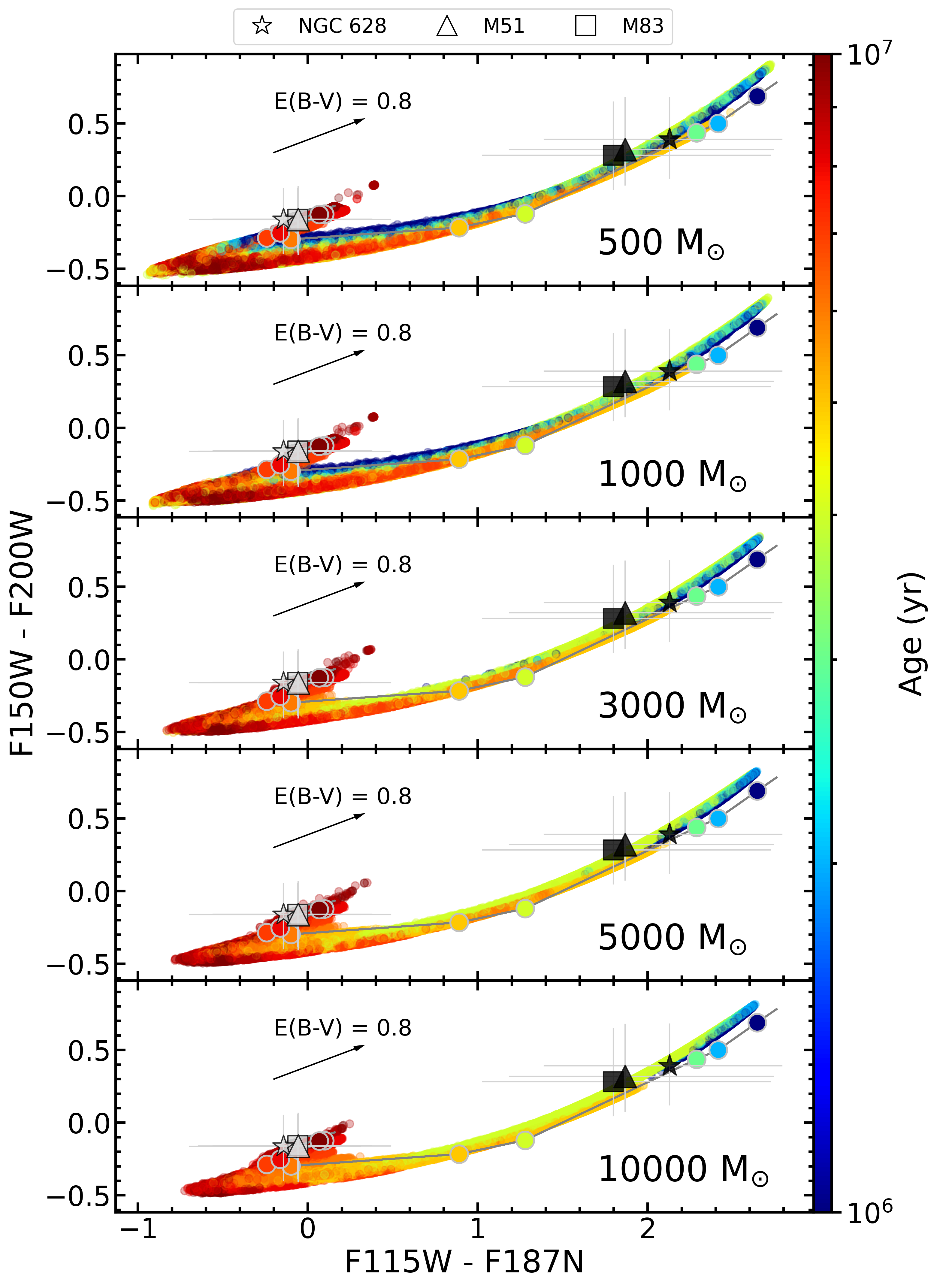}
    \caption{NIR color-color diagrams in ABmag for simulated \texttt{slug} cluster libraries with different fixed stellar mass and color-coded by cluster age.  The grey line represents a stellar evolutionary track computed with the {\tt yggdrasil} code. Each point of the track is color-coded by age as the \texttt{slug} cluster libraries. For an observational comparison, we plotted median values and 16th-84th percentiles for the young cluster population in M51, NGC~628 and M83. We show the eYSCI populations with black symbols, while gray symbols represent the optically identified YSC populations younger than 10 Myr and with a fitted E(B-V) $<0.1$, for a better comparison with the dust-free libraries. The extinction law vectors are from \cite{Fahrion23}. }
    \label{fig:slug_noEM}
\end{figure*}
We focus on the NIR bands and we plot the photometry of the simulated \texttt{slug} cluster libraries for solar metallicity in Fig.~\ref{fig:slug_noEM} (the equivalent figure for 40\%  solar metallicity models is included in Appendix~\ref{app:slug}). While in the left panel we show the color-color distribution from the NIR broad band filters that show the excess discussed in Sec.~\ref{sec:NIRexcess}, in the right panel we added the narrow band filter F187N, tracer of Pa$\alpha$ emission and therefore of cluster ages, to better compare our observations with the models. 
In this plot, for a specific value of mass, the simulated libraries of star clusters are represented by filled circles color coded by the age. In addition, we plot an evolutionary track from the {\tt yggdrasil} code (grey line) where each bin of age is color coded as the cluster library. The track is modeled for a 1000 M$_{\odot}$ cluster with solar metallicity, $f_{\rm cov} = 0.5$ and E(B-V) $= 0$. Similar to CIGALE, the {\tt yggdrasil} tracks are computed using a deterministic approach and are shown here for comparison against the {\tt slug} libraries.
Finally, we report median values of observed colors in M51, NGC~628 and M83 (for NGC~4449, see Appendix~\ref{app:slug})  for eYSCI (black symbols) and optical identified clusters younger than 10 Myr and with fitted E(B-V)$<0.1$ (gray symbols, see Sec.~\ref{sec:opticalCat}). 

In the left panel of Fig.~\ref{fig:slug_noEM}, we note that modeling emission from clusters using a stochastic approach can strongly impact the color differences of the NIR bands for low mass and young clusters. Focusing on a specific bin of younger ages ($< 5-6$ Myr), the recovered scatter in the NIR color-color distributions significantly increases going towards lower mass bins. For instance, at 1 Myr (blue circles) the scatter F115W - F150W in the x-axis goes from $\sim 0.5$ mag at 500 M$_{\odot}$ to $\sim 0.1$ mag at 10000 M$_{\odot}$. This effect also manifests as a scatter around the {\tt yggdrasil} tracks. For older clusters ($7-10$ Myr, orange to red circles), the spread becomes less evident as the NIR bands start being dominated by the presence of more evolved stars.  In the right panels of Fig.~\ref{fig:slug_noEM}, the inclusion of the narrow band filter transmitting Pa$\alpha$ scatters the color F115W$-$F187N over more then 2 magnitudes at very young ages, especially for low mass clusters ($<1000$ M$_{\odot}$). Even in this color-color space the direction of reddening, stochastic sampling and age coincide.

The median values recovered from observations in the four galaxies confirm that optically selected YSCs are on average older than eYSCs, with the median values settling where simulated clusters indicate ages from 6 to 10 Myr. We therefore expect that PMS stars have little contribution to their NIR colors. 

On the other hand, eYSCs are, on average, younger and their NIR colors might be partially explained by models that include both stochastic effects and attenuation as primary factors. The median value for fitted values of E(B-V) we presented in Sec.~\ref{sec:CIGres} indicates a reddening for eYSCs that is consistent with the reddening vector plotted in the figure, suggesting that both attenuation and stochastic sampling would move the models closer to the median colors of the eYSCs. The right panel of Fig.~\ref{fig:slug_noEM} indicates that the median ages of these sources in these galaxies are at face value close to 4-5 Myr old. However, the correction for extinction would make them younger  than the median values obtained with CIGALE.  

This analysis efficiently demonstrated that a stochastic approach, as well as the inclusion of PMS stars in the SED modeling process, might be important in understanding the NIR emission coming from low mass eYSCs. For instance, in the age range from 1 to 3 Myr and for stellar masses of 500-1000 M$_{\odot}$, 
the average scatter between \texttt{slug} and the {\tt yggdrasil} track in the F115W-F150W color is approximately 0.35 mag. The scatter observed in Fig.~\ref{fig:NIRexcess} for the low mass bin and the same range of ages is about 0.6 mag for the spiral galaxies and about 0.4 for NGC~4449. While the NIR excess observed for NGC~4449 makes stochastic IMF sampling combined with extinction a more likely explanation, for the spiral (metal-rich) environments it suggests that this effect alone might not be sufficient to explain the observed discrepancy. We will further discuss the observed differences between spirals and NGC~4449 in the next section.

An additional caveat to consider when quantifying the observed NIR excess using \texttt{slug} concerns the modeling of PMS stars within SSP synthesis codes. For low-mass stars, the choice of the initial stellar radius sets the internal entropy content, which in turn determines the stellar properties during the first $\sim 5$ Myr \citep{Hosokawa11,Haemmerle19}. Furthermore, observations have revealed that PMS stars undergo a dynamic accretion history, leading to a spread in entropy at birth \citep{Hunter17,Jensen18}. For low-mass stars with low effective temperatures, this effect can be significant and may help explain why stochastic models still fail to fully reproduce the observed NIR scatter.

\subsection{Missing ingredients to reproduce the 1--to--5~$\mu$m SED of eYSC}
\label{sec:missingIngredients}

Stellar evolutionary models included in \texttt{slug} account for PMS stars, but do not include emission from hot dust coming from their inner circumstellar disk and/or envelope surrounding accreting stellar cores. The spectra of a very young star \citep[young stellar object (YSO), e.g.,][]{Gutermuth11} surrounded by a dusty disk \citep[proto-planetary disk, e.g.,][]{Lada87} presents prominent emission in NIR wavelengths that has been addressed as "NIR excess" \citep{Hillenbrand98}, displaying redder colors when compared to only stellar emission. YSO SEDs will also contribute to the 3--to--5 $\mu$m. Thus, their inclusion might also alleviate the unrealistic dust grid models that the fit is currently preferring as presented in Section~\ref{sec:CIGgrid}.

In general, strong radiation fields may contribute to the dissolution of dusty disks; various studies have showed that proto-planetary disks can survive in regions of high and embedded star formation \citep[e.g.,][]{Richert15,Richert18}. Moreover, in a recent review, \cite{Winter22} illustrated that disks in low mass star forming regions are in average more likely to survive than the higher mass counterpart. The fact that the observed infrared excess is strongest for the lower-mass eYSCI could indicate that the contribution of YSOs might be an ingredient to consider in interpreting the SEDs of star clusters. This could be further tested with the inclusion of these additional models in SED fitting codes.

Furthermore, we noted above that the recovered residuals in NGC~4449 are smaller. Various studies have shown that in low metallicity environments proto-planetary disks might have less shielding from dust and therefore go towards a faster photo-evaporation \citep[e.g.,][]{Yasui10,Matsukoba24}. On the other hand, \cite{Demarchi24} investigated this issue using JWST NIRSpec observations of a sample of candidates accreting PMS stars. The authors concluded that PMS stars in lower metallicity environments (1/8 Z$_{\odot}$) efficiently maintain proto-planetary disks longer than solar metallicity counterparts, challenging the previous interpretation. 

While a contribution from YSOs may be crucial to explain the NIR excess observed in this work, a combination of additional phenomena must also be considered for a comprehensive understanding.

For example, we note that our SED fitting methodology does not account for the clumpiness of the dust distribution. Non-isotropic dust geometries can have a non-negligible impact on the NIR SEDs of clusters and on the resulting mass estimates \citep[e.g.,][]{Indebetouw06,Whelan11}.

Binary interactions and rotation-driven processes can also inject NIR light into the SED of a young star cluster. For instance, classical Be stars are fast rotators surrounded by a dust-free gaseous disk that emits significantly in the NIR \citep{Meilland12}. In addition, Wolf–Rayet stars, which are the classical examples of hydrogen‐stripped cores, are known to exhibit NIR excesses, as characterized by several previous IR surveys \citep[e.g.,][]{Johnson04,Crowther06,Mauerhan09,Mauerhan11}.

Finally, we note that our results rely on the assumption that each object in our sample is a single star cluster formed during a single burst of star formation (see Sec.~\ref{sec:CIGgrid}) and can therefore be described by a SSP model. In reality, we acknowledge that some clusters may instead consist of multiple sub-clusters, potentially leading to an age spread of a few Myr.



\subsection{Impact on physical properties} \label{sec:EW}


We investigated how the observed discrepancy between models and NIR observations affects the age estimates of eYSCs. To this end, we independently re-derived cluster ages using Pa$\alpha$ equivalent width (EW) measurements.

We estimated Pa$\alpha$ EWs using our photometric data as:
\begin{equation}
    {EW}_{{\rm Pa}\alpha} = \frac{F_{\lambda,{\rm line}}}{F_{\lambda,{\rm cont}}}{BW}_{\rm F187N}
\end{equation}
with $F_{\lambda,{\rm line}} = F_{\lambda,{\rm tot}} - F_{\lambda,{\rm cont}}$ being the intrinsic flux density of the Pa$\alpha$ emission line, $F_{\lambda,{\rm tot}}$ the total flux density measured in the F187N filter and $F_{\lambda,{\rm cont}}$ the measured underlying continuum resulting from the interpolation of a blue component (F150W) and a red one (F200W), using the same reasoning for continuum subtraction presented in Sec.~\ref{sec:eyscPresentation}. $BW_{\rm F187N}$ corresponds to the nominal band width of the F187N filter: 240 {\AA} \footnote{\url{https://jwst-docs.stsci.edu/jwst-near-infrared-camera/nircam-instrumentation/nircam-filters\#gsc.tab=0}}. We did not apply any extinction as the Pa$\alpha$ EW is only marginally affected by it \citep{Messa21}. To estimate eYSCI ages we interpolated the Pa$\alpha$ EWs obtained from our samples of eYSCI in M51, M83, NGC~628 and NGC~4449, with recovered EWs from the {\tt yggdrasil} tracks introduced in Sec.~\ref{sec:slug}. In the case of NGC~4449, we used a stellar track with 40\% solar metallicity.
\begin{figure*}[htb]
    \centering
    \includegraphics[width = 0.9\textwidth]{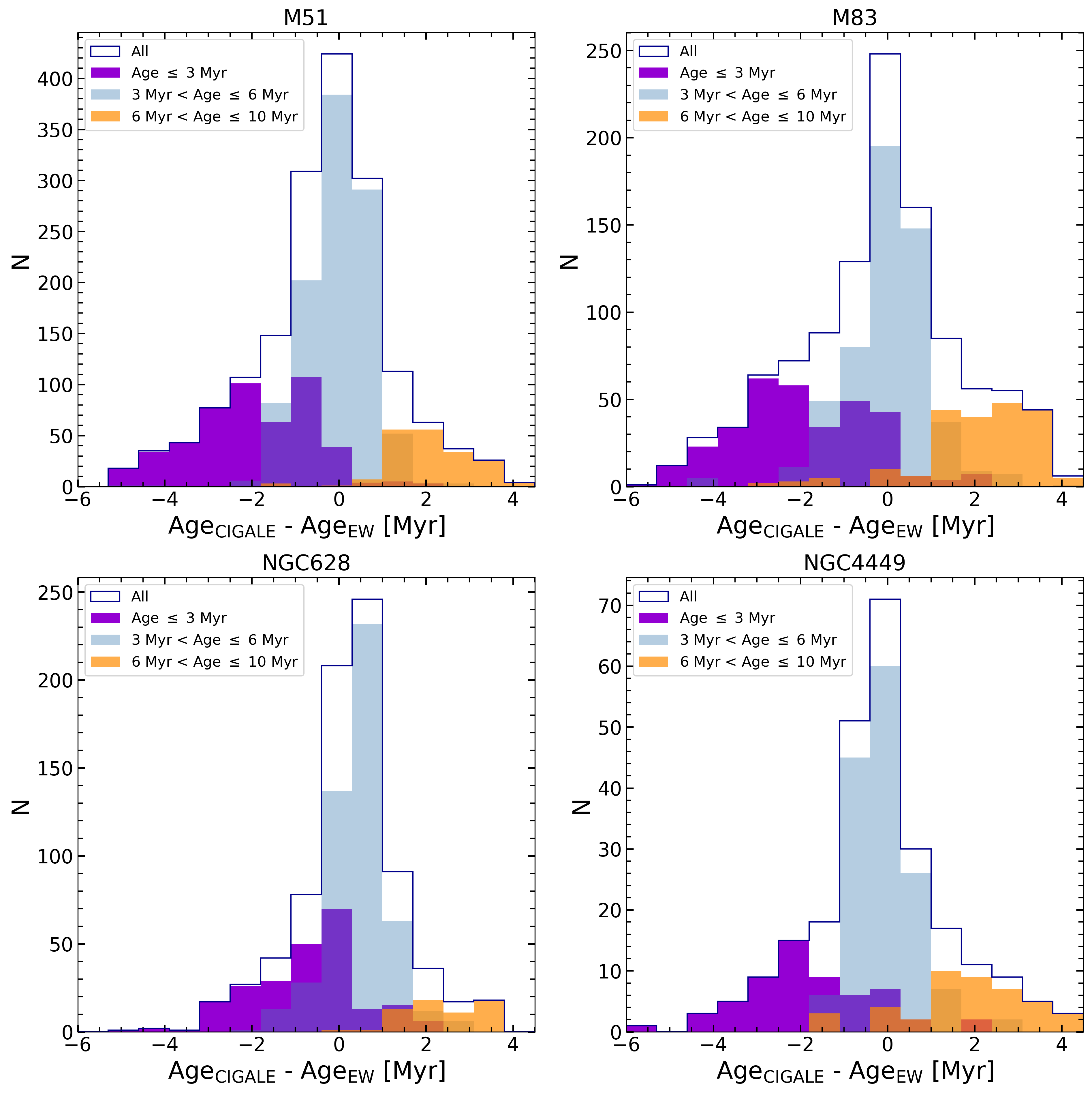}
    \caption{Distribution of age differences for eYSCI in M51, M83, NGC~628 and NGC~4449, obtained using two different methods. For each panel, on the x-axis, Age$_{\rm CIGALE}$ represents the best fit values of age from CIGALE, while Age$_{\rm EW}$ is estimated from Pa$\alpha$ EWs, as described in Sec.~\ref{sec:EW}. The unfilled histogram shows the entire population of eYSCI in the four FEAST galaxies, while we use filled histograms to compare the age differences using as a reference the CIGALE bin of ages defined in Sec.~\ref{sec:NIRexcess}. Age bins are color-coded as in Fig.~\ref{fig:NIRexcess}.}
    \label{fig:EW_Age}
\end{figure*}
We compare the recovered eYSCI ages from CIGALE and the EW method in Fig.~\ref{fig:EW_Age}.
The blue unfilled distribution shows the age difference for the entire eYSCI populations in the four FEAST galaxies, while filled histograms correspond to the distributions binned by the CIGALE best fitted ages (see Sec.~\ref{sec:NIRexcess}) and color-coded as Fig.~\ref{fig:NIRexcess}. 

Overall, we find that age differences between the CIGALE and EW methods are, on average, within $\sim$1 Myr in 80\% of the cluster population analyzed in this work. However, we also recovered larger deviations. The subsample of eYSCs with ages older than 6 Myr in the CIGALE fit (orange colors in Fig.~\ref{fig:NIRexcess}) shows EWs consistent with significantly younger ages, in agreement with the fact that they are still associated to considerable ionized H emission. The subset of eYSCI with CIGALE best ages $<3$ Myr (purple) shows older EW ages. These trends may result from different underlying causes. 

It is reasonable to assume that the EW estimates might be affected by the NIR excess in the continuum underlying Pa$\alpha$. CIGALE and other evolutionary models like {\tt yggdrasil} do not account for additional components in the NIR, and consequently, the interpolated observed EWs might be underestimated, biasing the age distribution toward older values. 
This effect could be major in a regime where the contribution to the observed flux from the source of the excess is higher than the continuum produced by the stellar population models. Based on the measured excess in the $<3$ Myr subsample (purple distributions), this impact is not negligible for $\sim 30-40$\% of the objects, potentially leading to lower EWs and thus older inferred ages. 

\begin{figure}[htb]
    \centering
    \includegraphics[width = 0.9\columnwidth]{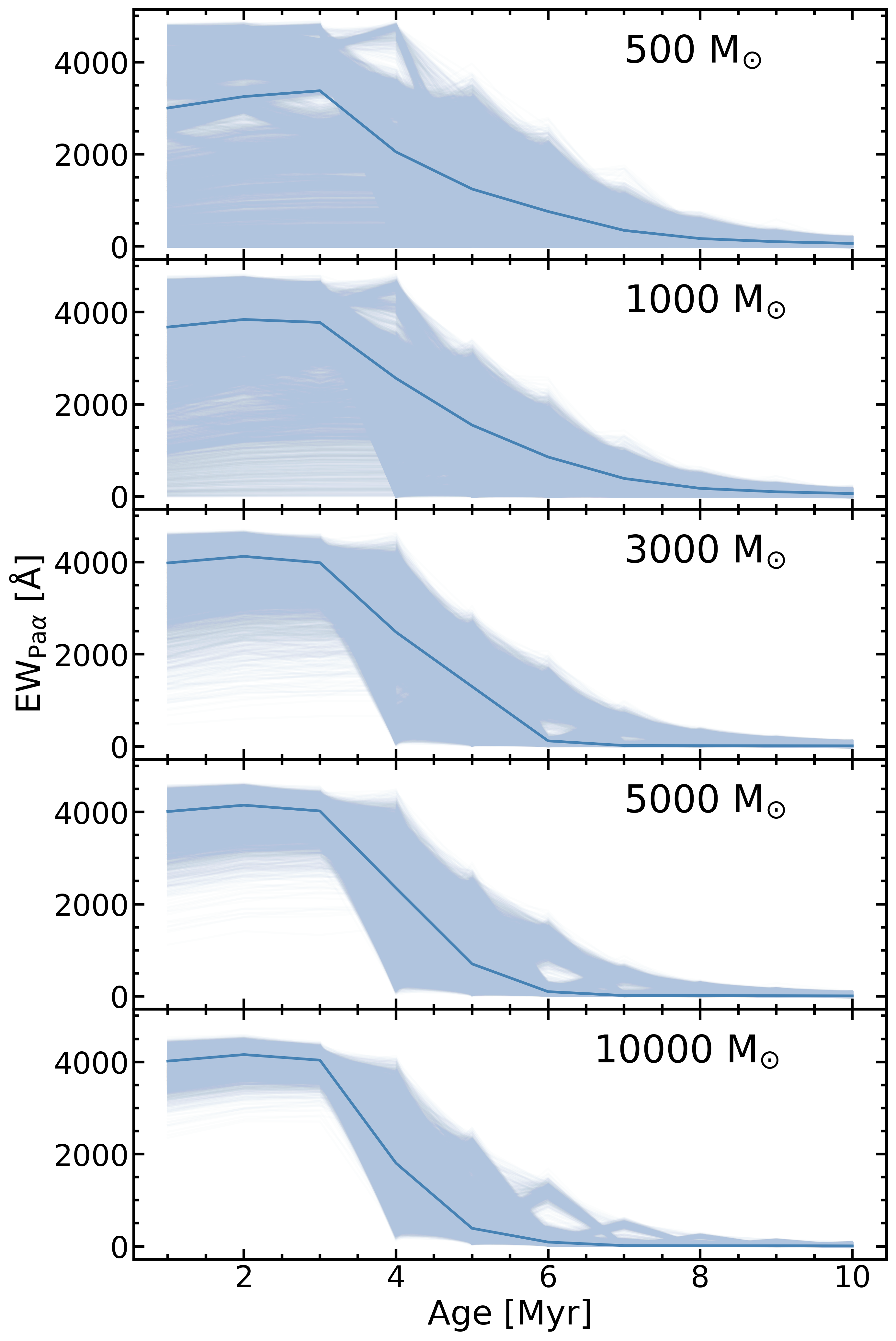}
    \caption{Pa$\alpha$ EW as a function of age for simulated {\tt slug} cluster libraries with different fixed stellar masses of 500, 1000, 3000 5000 and 10000 M$_{\odot}$, as defined in Sec.~\ref{sec:slug}. Each panel corresponds to a specific stellar mass, as indicated by the label. The darker lines represent the median tracks of the simulated populations.}
    \label{fig:EW_slug}
\end{figure}

Finally, stochastic IMF sampling also affects the EW-age relation commonly used in the literature. In Fig.~\ref{fig:EW_slug}, we present the recovered Pa$\alpha$ EWs from our {\tt slug} simulations (see Sec.~\ref{sec:slug}) as a function of age for different cluster masses. For each mass bin, the darker lines indicate the median values. We find significant differences across cluster masses, both in individual realizations and in the median trends. In the low mass bins, the EW-age relation breaks down: for example, a low observed EW could correspond to either a young or an old cluster. The relation only emerges for cluster masses larger than 5000 M$_{\odot}$. These trends implies that even EW measurements cannot be trusted as age indicators.


Our analysis suggests that caution must be taken when interpreting results that rely on inferred physical properties of eYSCs. 

\section{Conclusions} \label{sec:conclusion}

In this paper, we used HST and JWST/NIRCam observations to analyze the SEDs of the population of eYSCI in four galaxies from the FEAST survey: M51, M83, NGC~628 and NGC~4449. These clusters are selected to be well detected in ionized hydrogen emission (Pa$\alpha$) and to exhibit a prominent 3.3 $\mu$m PAH feature from their surrounding PDRs, indicating that they are very young ($< 10$ Myr) and possibly dusty. After thoroughly exploring the fitting parameter space (see Sec.~\ref{sec:CIGgrid}), we modeled their SEDs using the CIGALE fitting code. The main findings of this analysis are summarized as follows:

\begin{itemize}

    \item The dust parameter grid that yields better fits with CIGALE deviates from expectations based on MIR–FIR studies of star-forming regions. The adopted dust grid in this work is not physically motivated, but rather chosen solely based on the quality of the fits.
    
    \item We identify a NIR flux excess that cannot be fully reproduced by current models. We find that the CIGALE fits in the F150W and F200W filters reveal a systematic discrepancy, with the best-fit fluxes underestimating the observed values. We investigated whether observational biases or the choice of model parameters could resolve this discrepancy, and conclude that they cannot account for the observed excess. This excess may originate from one or more stellar or hot dust components in the SED that are not accounted for in existing models.

    \item The observed NIR excess correlates with the ages and stellar masses of the eYSCs. Young (age $\leq 3$ Myr) and low-mass (M$_\star \leq 3000$ M$_{\odot}$) clusters exhibit the strongest positive residuals of about 0.3-0.6 mag in F150W and F200W. Clusters with fitted ages $> 6$ Myr do not show a prominent broadband excess; however, high residuals in the narrowbands F187N and F405N, tracing ionized gas, suggest that these sources may in fact be younger than inferred from the best fit.

    \item We explored the effects of stochastic IMF sampling using the {\tt slug} code. Our results show that both stochasticity and the inclusion of PMS stellar emission in the models might partially account for the observed color excess and consequently the unconstrained dust grids. We further speculate that including emission from YSOs may be essential to reduce these discrepancies.

    \item We investigated the impact of the observed NIR excess on age estimates. We derive the cluster ages using color-color distributions, SED fits and EW estimates.These different methods point toward very young star cluster populations, however, none of them yield reliable absolute age determinations.
    
\end{itemize}

Our results highlight the limitations of current SED fitting tools like CIGALE in modeling eYSC SEDs, particularly in reproducing the observed NIR properties. The NIR excess is most pronounced in very young ($\leq 3$ Myr), low-mass systems. In these regimes, stochastic effects, PMS star evolution, YSO disk emission, and other additional contributions discussed in the text may become increasingly important, yet are not fully captured in standard models. Ideally, future tools would incorporate stochastic sampling across all stellar evolutionary phases and include more realistic treatments of both stellar and dust emission in young clusters. To minimize systematics due to modeling assumptions, such tools would benefit from a design in which different components of the models can be turned on or off, allowing for a more controlled exploration of their individual effects on the resulting SEDs.


\section*{acknowledgments}
The authors thank the anonymous referee for their valuable comments, which have greatly improved the quality of this paper.
This work is based in part on observations made with the NASA/ESA/CSA James Webb Space Telescope, which is operated by the Association of Universities for Research in Astronomy, Inc., under NASA contract NAS 5-03127. The data were obtained from the Mikulski Archive for Space Telescopes (MAST) at the Space Telescope Science Institute. These observations are associated with program \# 1783. Support for program \# 1783 was provided by NASA through a grant from the Space Telescope Science Institute, which is operated by the Association of Universities for Research in Astronomy, Inc., under NASA contract NAS 5-03127. The specific observations analyzed can be accessed via\dataset[10.17909/f4vm-c771]{http://dx.doi.org/10.17909/f4vm-c771}. All our data products are available at MAST as a High Level Science Product via https://doi.org/10.17909/6dc1-9h53 and DOI: 10.17909/6dc1-9h53 and on the FEAST webpage https://feast-survey.github.io/.
A.A and A.P. acknowledge support from the Swedish National Space Agency (SNSA) through the grant 2021- 00108. A.A. and H.F.V. acknowledges support from SNSA 2023-00260.
KG is supported by the Australian Research Council through the Discovery Early Career Researcher Award (DECRA) Fellowship (project number DE220100766) funded by the Australian Government. 
KG is supported by the Australian Research Council Centre of Excellence for All Sky Astrophysics in 3 Dimensions (ASTRO~3D), through project number CE170100013. ADC and ASMB acknowledge the support from the Royal Society University Research Fellowship URF/R1/191609 (PI: ADC). MM acknowledges financial support through grants PRIN-MIUR 2020SKSTHZ, the INAF GO Grant 2022 “The revolution is around the corner: JWST will probe globular cluster precursors and Population III stellar clusters at cosmic dawn,” and by the European Union – NextGenerationEU within PRIN 2022 project n.20229YBSAN - "Globular clusters in cosmological simulations and lensed fields: from their birth to the present epoch”.

%






\appendix

\section{Comparison of SED fitting with exclusion of H emission lines}\label{app:opt4449}
\begin{figure*}[htb]
    \centering
    \includegraphics[width = 0.46\textwidth]{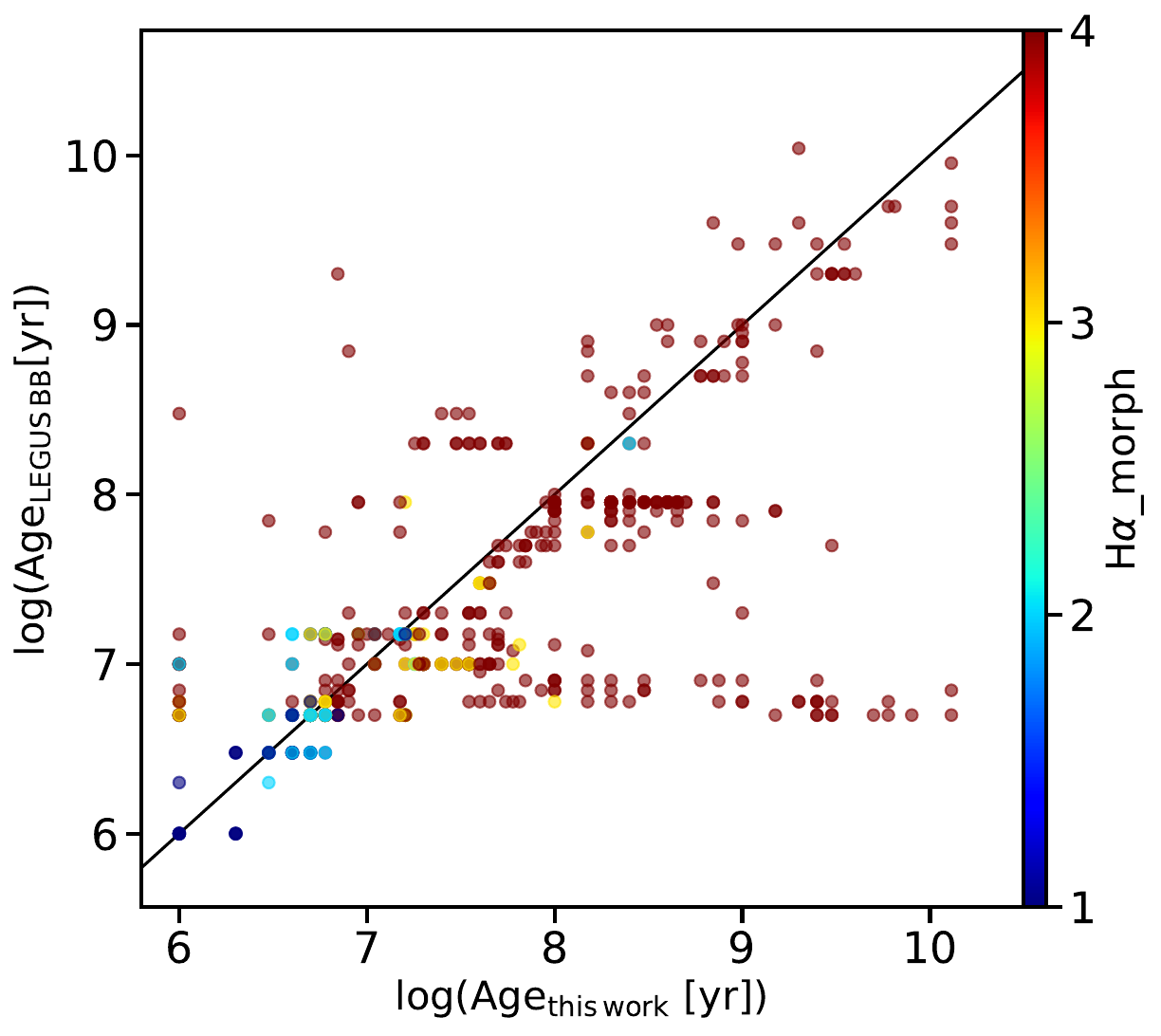}
    \hfill
    \includegraphics[width = 0.46\textwidth]{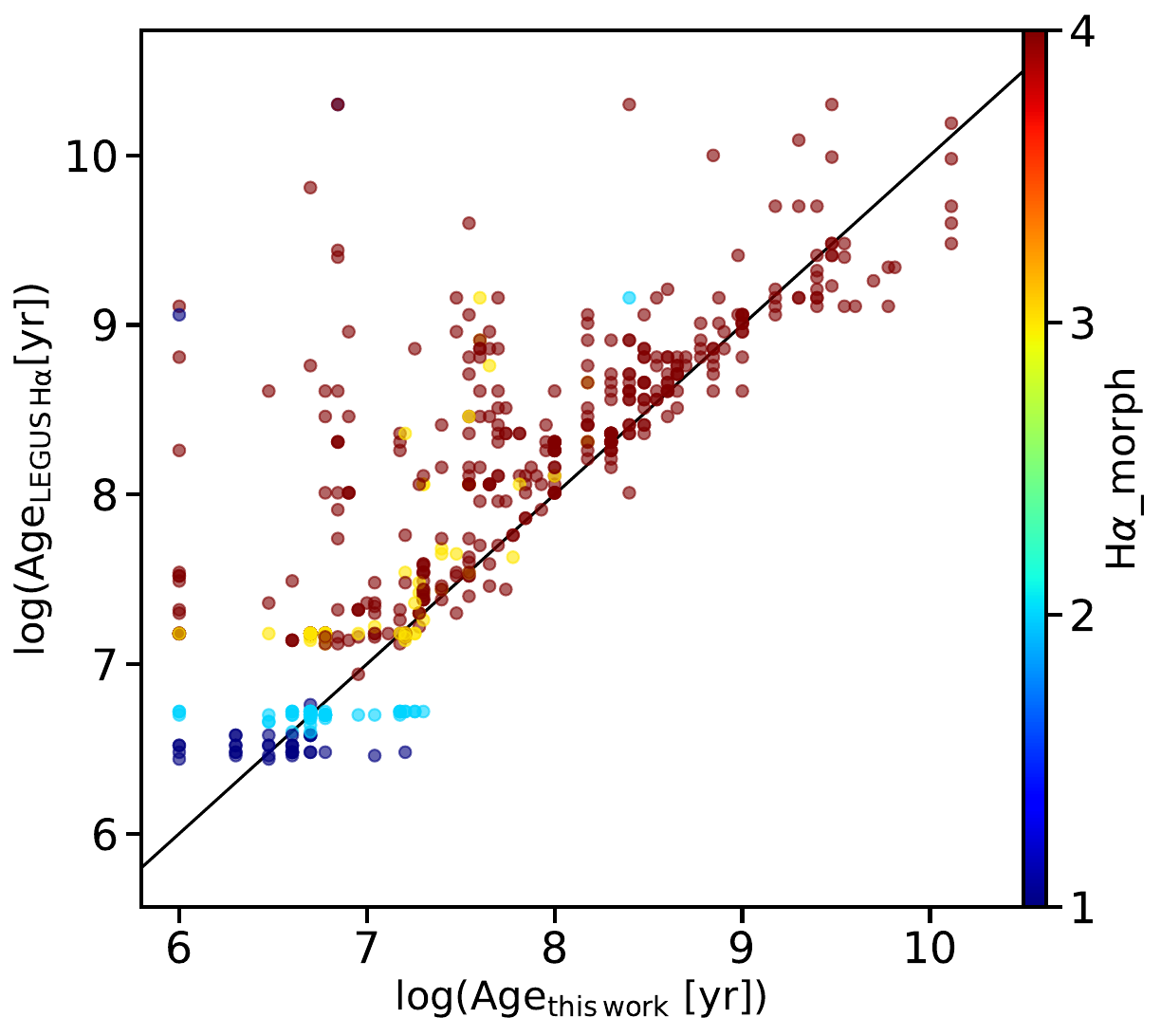}
    \caption{Comparison between the CIGALE-estimated ages of the optical YSC population in NGC~4449 (see Sec.~\ref{sec:opticalCat}) and the age estimates from the LEGUS survey \citep{Whitmore20}. In the left panel, LEGUS ages are derived using only broadband filters, while in the right panel, the H$\alpha$ recombination line is included. The color bar indicates the H$\alpha$ morphology (from 1: strong H$\alpha$ emission, to 4: no H$\alpha$), as described in \citet{Whitmore20}.}
    \label{fig:NGC4449_SEDfits}
\end{figure*}

This section of the appendix presents a comparison between our age estimates for optically selected YSCs in NGC~4449 obtained with CIGALE, as described in the main text, and previous results from the LEGUS survey \citep{Calzetti15}. The goal of this test is to highlight the importance of including hydrogen recombination lines in the SED fitting of YSCs. In the left panel of Fig.~\ref{fig:NGC4449_SEDfits}, we compare our age estimates with those derived using only HST broadbands \citep[F275W, F336W, F435W, F555W, F814W;][]{Whitmore20}. In the right panel, by contrast, the LEGUS ages are derived with the addition of the H$\alpha$ emission line \citep[F658N;][]{Whitmore20}. As this comparison shows, the inclusion of even a single hydrogen recombination line in the SED significantly improves the fit quality, helping to mitigate the age–extinction degeneracy.

\section{Aperture effects in M83}\label{app:aperture}

\begin{figure*}[htb]
    \centering
    \includegraphics[width = 0.46\textwidth]{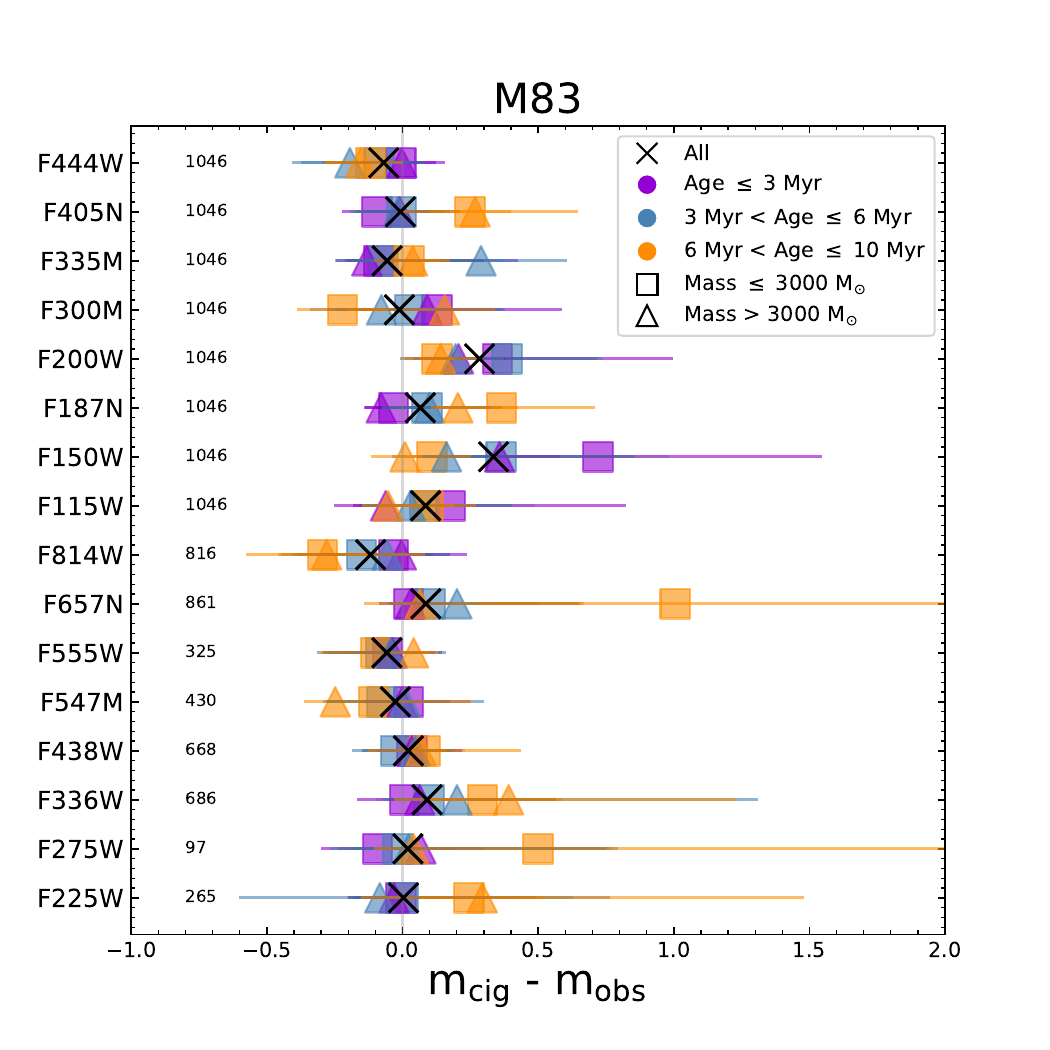}
    \hfill
    \includegraphics[width = 0.46\textwidth]{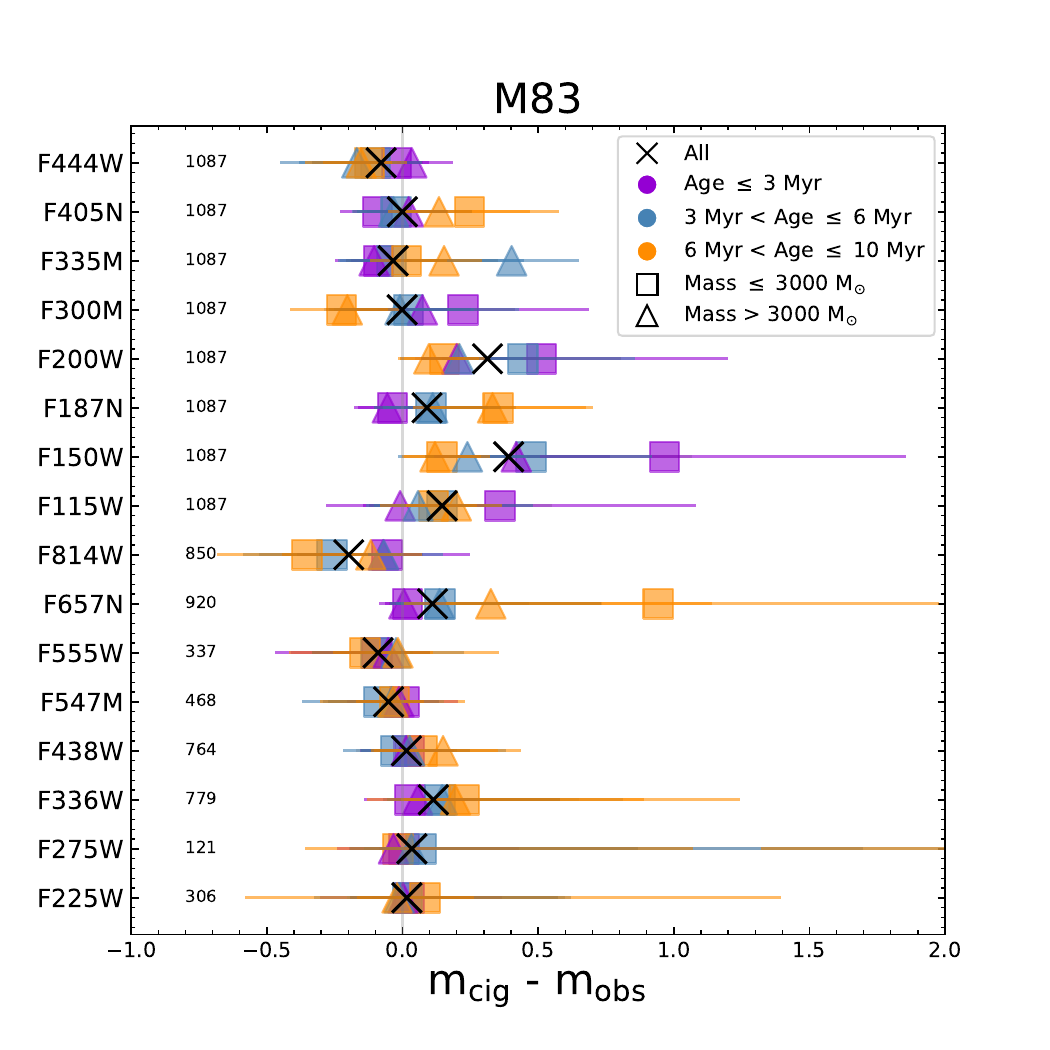}
    \caption{Same as the top right panel of Fig.~\ref{fig:NIRexcess}, but adopting radii of 4 ($\sim$ 3.6 pc, left) and 6 pixels ($\sim$ 5.5 pc, right) for the aperture photometry measurements.}
    \label{fig:M83_apertures}
\end{figure*}

As introduced in Sec.\ref{sec:discussion}, we tested different photometric aperture sizes to assess the impact on the observed NIR excess. Here, we present this analysis for one of the FEAST targets: M83. While the photometric radius adopted in the main analysis is 5 pixels, Fig.\ref{fig:M83_apertures} shows the corresponding residual analysis for apertures of 4 pixels (left panel) and 6 pixels (right panel), analogous to the one shown in the top right panel of Fig.~\ref{fig:NIRexcess}. As discussed in the text, we do not observe significant changes in the NIR excess when using apertures of 4 or 5 pixels. However, the residuals increase when a 6-pixel aperture is used.

\section{Residual analysis of the YSC populations in M83, NGC~628 and 4449}\label{app:optapp}
\begin{figure*}[htb]
    \centering
    \includegraphics[width = 0.31\textwidth]{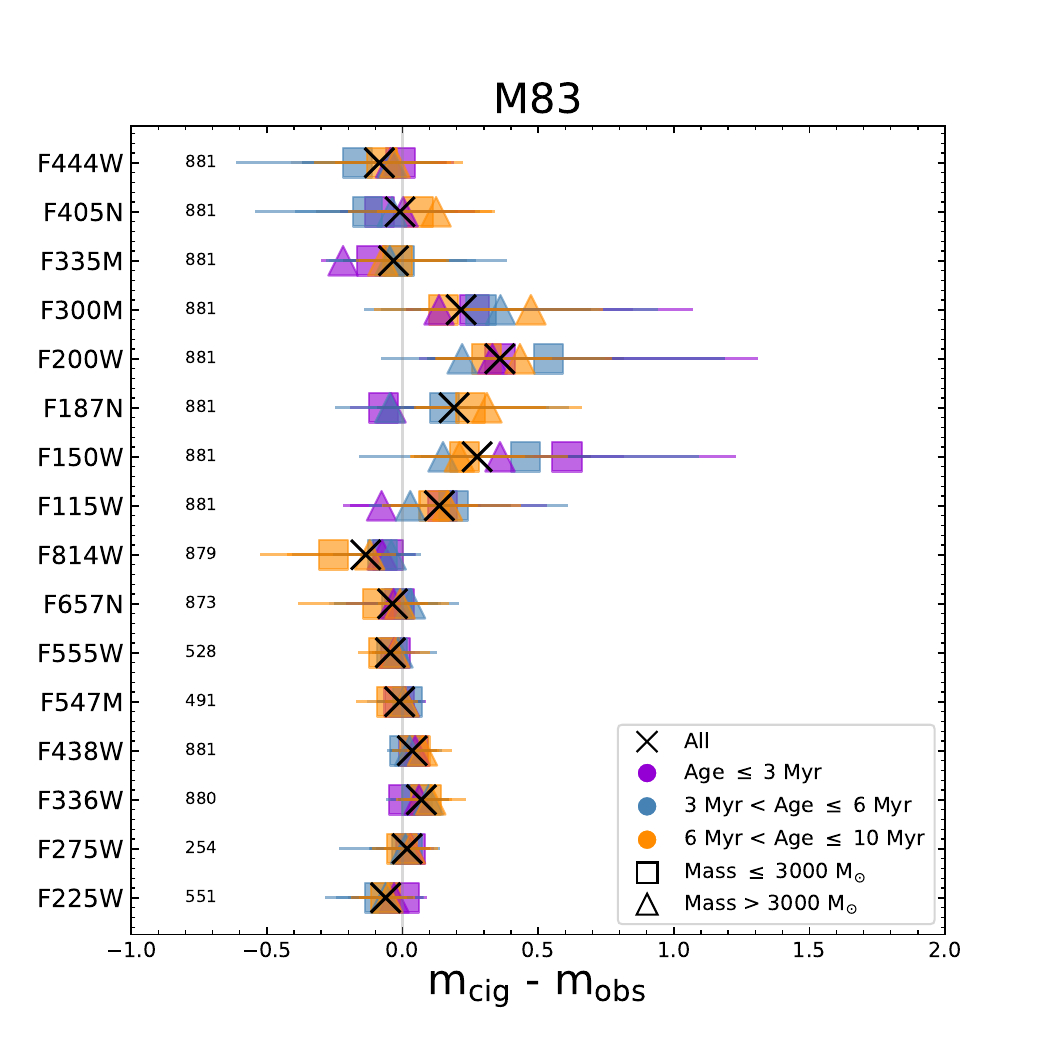}
    \hfill
    \includegraphics[width = 0.31\textwidth]{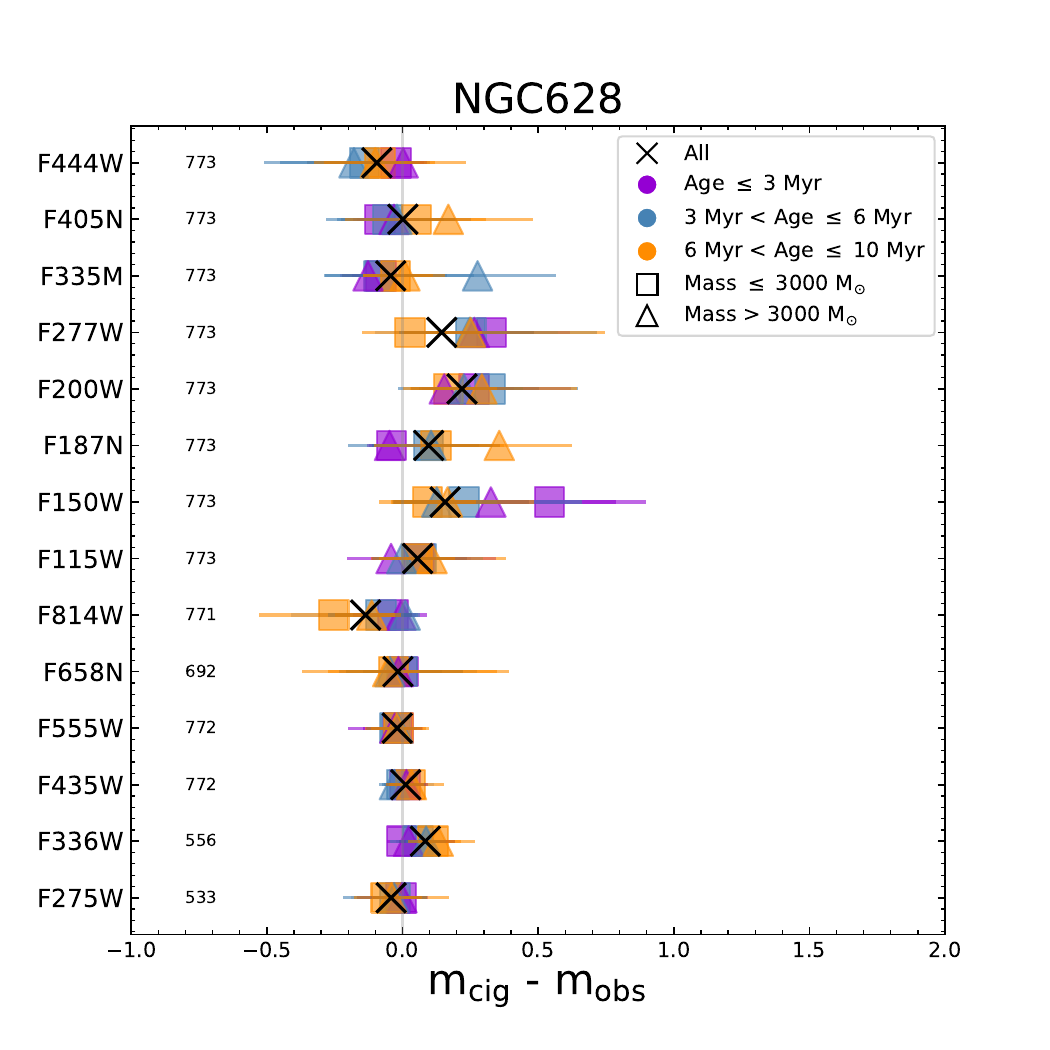}
    \hfill
    \includegraphics[width = 0.31\textwidth]{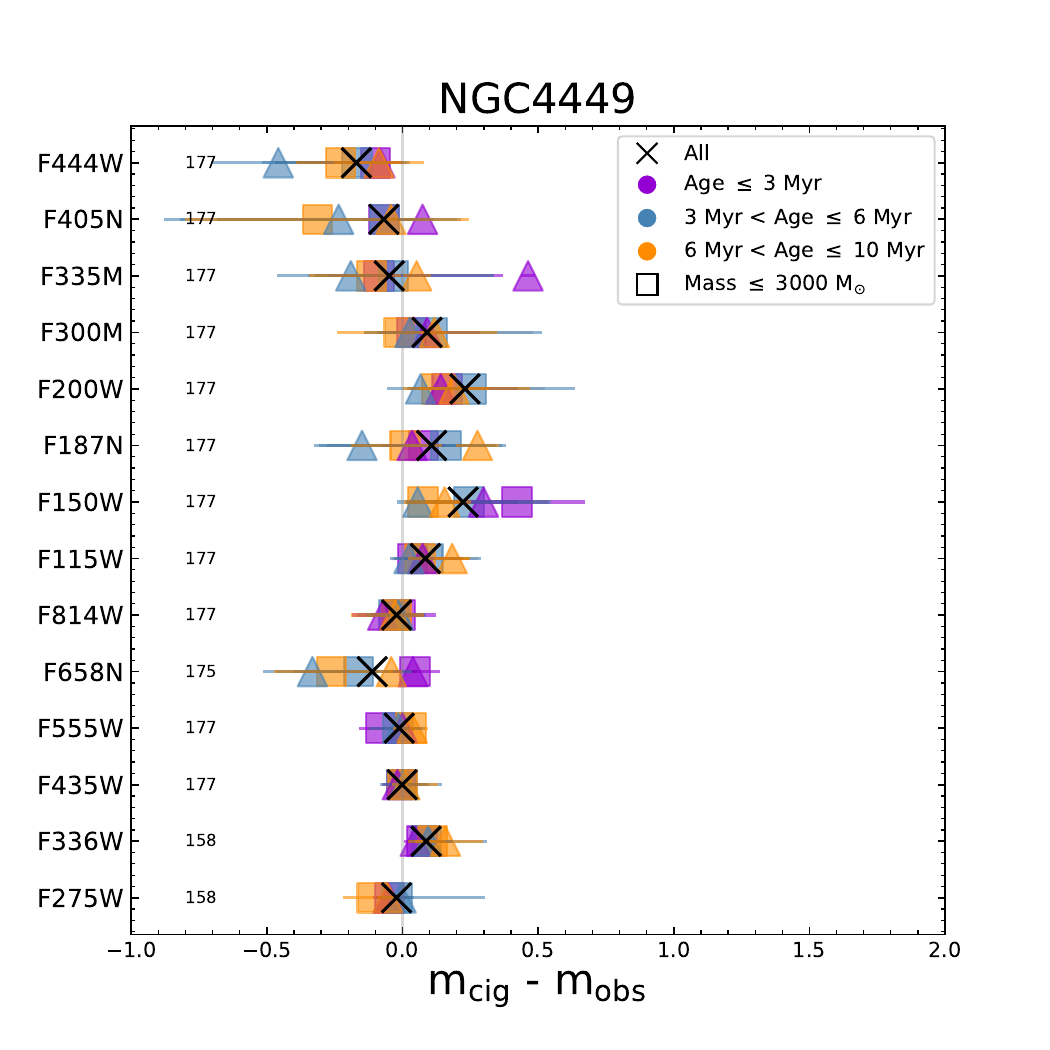}
    \caption{Same as Fig.~\ref{fig:res_OPT}, but for the YSC populations in M83 (left), NGC~628 (center), NGC~4449 (right).}
    \label{fig:YSC_opts}
\end{figure*}

In Sec.~\ref{sec:dustDisc}, we discussed the presence of the NIR excess in the sample of optically selected YSCs in M51, probing that the excess is less evident for these objects compared to the eYSCI population (see Fig.~\ref{fig:res_OPT}). Here, we extend the same analysis to the other three galaxies. The results, shown in Fig.~\ref{fig:YSC_opts}, confirm the trend observed for M51.

\section{SLUG libraries for 0.4 solar metallicity}\label{app:slug}

With a metallicity of $12 + \log ({\rm O}/{\rm H}) = 8.26$ and an irregular morphology, NGC~4449 differs from the other three spiral galaxies analyzed in this paper. Consequently, the {\tt slug} analysis presented in Sec.\ref{sec:slug} (Fig.\ref{fig:slug_noEM}) requires a slightly modified approach. To account for this, we generate simulated {\tt slug} cluster libraries using stellar and nebular models with a fixed metallicity of 0.4$Z_\odot$.

\begin{figure*}[htb]
    \centering
    \includegraphics[width = 0.496\textwidth]{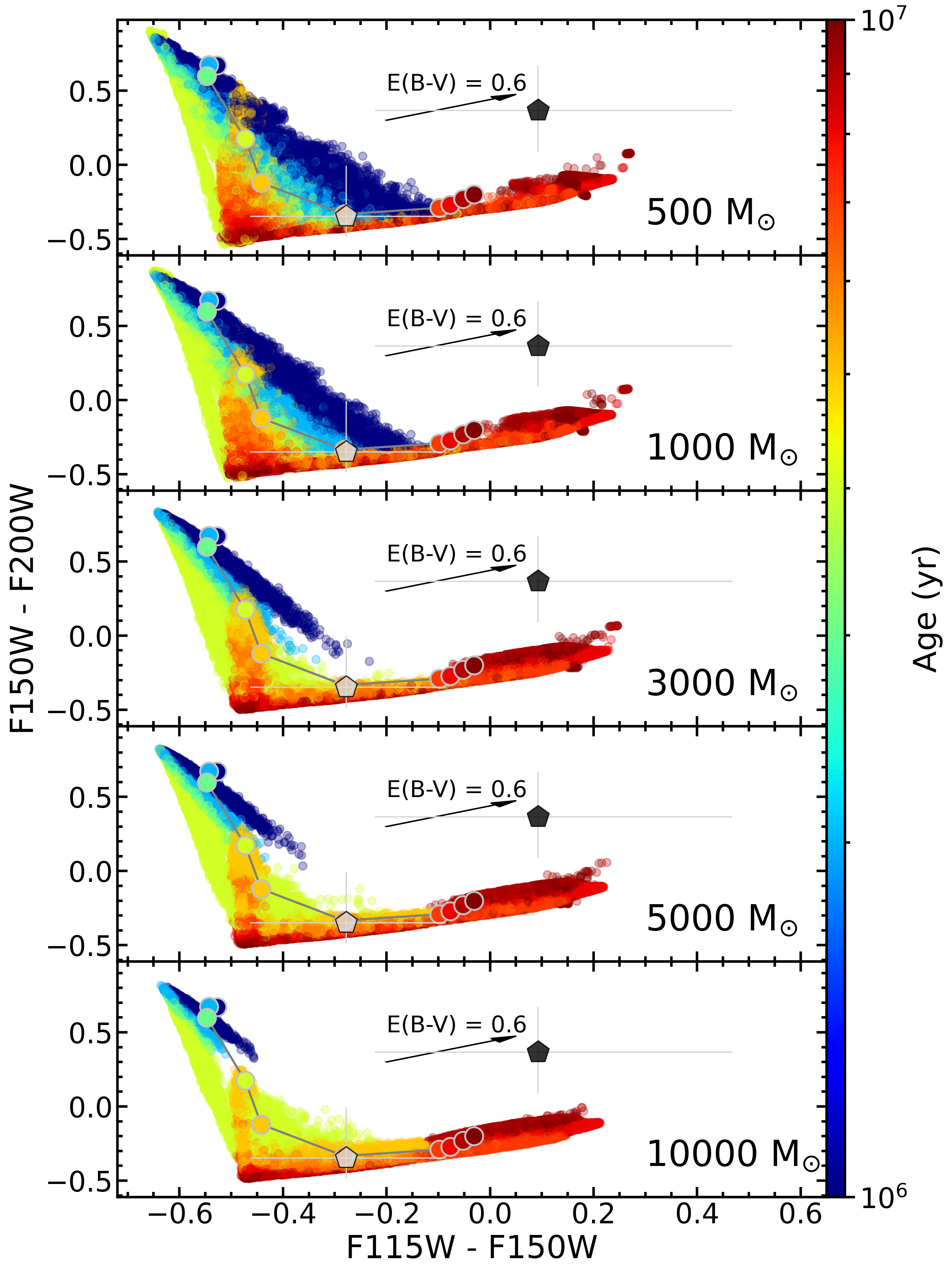}
    \hfill
    \includegraphics[width = 0.496\textwidth]{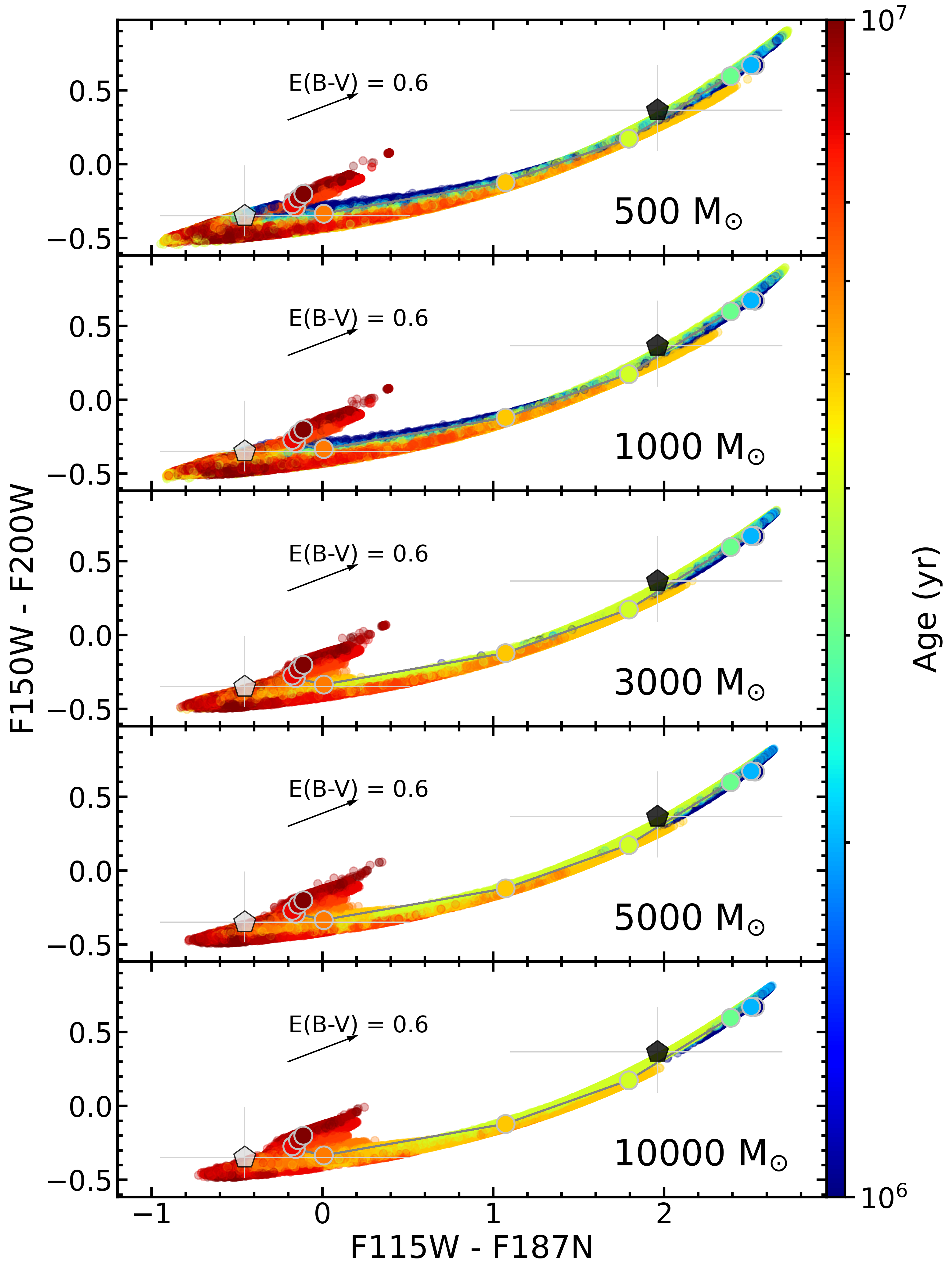}
    \caption{Same as Fig.~\ref{fig:slug_noEM}, but with adopted metallicity Z $= 0.4$ Z$_{\odot}$. Black and gray pentagons represent the eYSCI and optically identified YSC younger than 10 Myr and with a fitted E(B-V)$<0.1$ populations in NGC~4449, respectively.}
    \label{fig:slug_app}
\end{figure*}

The results, shown in Fig.\ref{fig:slug_app}, indicate that the color differences observed at this lower metallicity are consistent with those seen in Fig.\ref{fig:slug_noEM}.


\bibliography{Mypapers}{}
\bibliographystyle{aasjournal}



\end{document}